\documentclass[journal]{IEEEtran}
\IEEEoverridecommandlockouts

\usepackage{cite}
\usepackage{amsmath,amssymb,amsfonts}
\usepackage{amssymb}
\usepackage{algpseudocode}
\usepackage{amsfonts}
\usepackage{graphicx}
\usepackage{fancyhdr}
\usepackage{cases}
\usepackage{textcomp}
\usepackage{extarrows}
\usepackage{algorithm,algpseudocode}
\usepackage{orcidlink}

\allowdisplaybreaks

\usepackage{algorithm}
\usepackage{algpseudocode}

\usepackage{multirow}
\usepackage{caption}
\usepackage{float} 
\usepackage{subcaption}
\usepackage{makecell}
\usepackage{booktabs}

\usepackage{hyperref}
\usepackage{caption}

\usepackage{multirow,tabularx}
\usepackage{multicol,mwe,float,subcaption}
\usepackage{mathtools}
\usepackage{xcolor}
\usepackage[english]{babel}
\usepackage{amsthm}
\usepackage[numbers,sort&compress]{natbib}
\algnewcommand{\Inputs}[1]{%
  \State \textbf{Inputs:}
  \Statex \hspace*{\algorithmicindent}\parbox[t]{.8\linewidth}{\raggedright #1}
}
\algnewcommand{\Initialize}[1]{%
  \State \textbf{Initialization:}
  \Statex \hspace*{\algorithmicindent}\parbox[t]{.8\linewidth}{\raggedright #1}
}
\usepackage{caption}
\def\BibTeX{{\rm B\kern-.05em{\sc i\kern-.025em b}\kern-.08em
    T\kern-.1667em\lower.7ex\hbox{E}\kern-.125emX}}
    
\begin{document}


\title{\fontsize{23.8pt}{\baselineskip}\selectfont Robust Beamforming for RIS-aided Communications: Gradient-based Manifold Meta Learning}

	\author{
	\IEEEauthorblockN{
	Fenghao Zhu~\orcidlink{0009-0006-5585-7302},
	Xinquan Wang~\orcidlink{0009-0005-9986-7054},
	Chongwen Huang~\orcidlink{0000-0001-8398-8437},
	Zhaohui Yang~\orcidlink{0000-0002-4475-589X},
	Xiaoming Chen,
	Ahmed Alhammadi~\orcidlink{0009-0006-8535-268X},\\
	Zhaoyang Zhang~\orcidlink{0000-0003-2346-6228},~\IEEEmembership{Senior Member,~IEEE},
	Chau Yuen~\orcidlink{0000-0002-9307-2120},~\IEEEmembership{Fellow,~IEEE}, and M\'{e}rouane~Debbah~\orcidlink{0000-0001-8941-8080},~\IEEEmembership{Fellow,~IEEE}}

\thanks{
The work was supported by the China National Key R\&D Program under Grant 2021YFA1000500 and 2023YFB2904804, National Natural Science Foundation of China under Grant 62331023, 62101492, 62394292 and U20A20158, Zhejiang Provincial Natural Science Foundation of China under Grant LR22F010002, Zhejiang Provincial Science and Technology Plan Project under Grant 2024C01033, and Zhejiang University Global Partnership Fund, MOE Tier 2 (Award number MOE-T2EP50220-0019) and A*STAR (Agency for Science, Technology and Research) Singapore, under Grant No. M22L1b0110.

F. Zhu, C. Huang are with College of Information Science and Electronic Engineering, Zhejiang University, Hangzhou 310027, China, the State Key Laboratory of Integrated Service Networks, Xidian University, Xi’an 710071, China, and Zhejiang Provincial Key Laboratory of Info. Proc., Commun. \& Netw. (IPCAN), Hangzhou 310027 China (E-mails: \href{mailto:zjuzfh@zju.edu.cn}{\{zjuzfh}, \href{mailto:chongwenhuang@zju.edu.cn}{chongwenhuang\}@zju.edu.cn}).}

\thanks{
X. Wang, Z. Yang, X. Chen and Z. Zhang are with the College of Information Science and Electronic Engineering, Zhejiang University, Hangzhou 310027, China (E-mails: \href{mailto:wangxinquan@zju.edu.cn}{wangxinquan}, \href{mailto:yang_zhaohui@zju.edu.cn}{\{yang\_zhaohui}, \href{mailto:chen_xiaoming@zju.edu.cn}{chen\_xiaoming}, \href{mailto:ning_ming@zju.edu.cn}{ning\_ming\}@zju.edu.cn}).}

\thanks{A. Alhammadi is with Technology Innovation Institute, 9639 Masdar City, Abu Dhabi, UAE (E-mail: \href{mailto:Ahmed.Alhammadi@tii.ae}{Ahmed.Alhammadi@tii.ae}).}

\thanks{C. Yuen is with the School of Electrical and Electronics Engineering, Nanyang Technological University, Singapore 639798 (E-mail: \href{mailto:chau.yuen@ntu.edu.sg}{chau.yuen@ntu.edu.sg}).}

\thanks{M. Debbah is with KU 6G Research Center, Khalifa University of Science and Technology, P O Box 127788, Abu Dhabi, UAE (E-mail: \href{mailto:merouane.debbah@ku.ac.ae}{merouane.debbah@ku.ac.ae}).}



}

\maketitle

\pagestyle{empty}  
\thispagestyle{empty} 

\begin{abstract}
Reconfigurable intelligent surface (RIS) has become a promising technology to realize the programmable wireless environment via steering the incident signal in fully customizable ways. However, a major challenge in RIS-aided communication systems is the simultaneous design of the precoding matrix at the base station (BS) and the phase shifting matrix of the RIS elements. This is mainly attributed to the highly non-convex optimization space of variables at both the BS and the RIS, and the diversity of communication environments. Generally, traditional optimization methods for this problem suffer from the high complexity, while existing deep learning based methods are lacking in robustness in various scenarios. To address these issues, we introduce a gradient-based manifold meta learning method (GMML), which works without pre-training and has strong robustness for RIS-aided communications. Specifically, the proposed method fuses meta learning and manifold learning to improve the overall spectral efficiency, and reduce the overhead of the high-dimensional signal process. Unlike traditional deep learning based methods which directly take channel state information as input, GMML feeds the gradients of the precoding matrix and phase shifting matrix into neural networks. Coherently, we design a differential regulator to constrain the phase shifting matrix of the RIS. Numerical results show that the proposed GMML can improve the spectral efficiency by up to 7.31\%, and speed up the convergence by 23 times faster compared to traditional approaches. Moreover, they also demonstrate remarkable robustness and adaptability in dynamic settings.
\end{abstract}


\section{Introduction}\label{sec:intro}
Reconfigurable intelligent surface (RIS) is an artificial passive meta-surface that can reconstruct the wireless environment by manipulating the phase of incoming electromagnetic waves \cite{hcwRIS, hcwCE, wqqRIS, hcwRL}, and it is considered as a potential technology for the upcoming sixth-generation communication to improve the coverage and spectral efficiency (SE). Thanks to the passive phase shifting technology at the RIS and active beamforming technology at the base station (BS), both the throughput and energy efficiency of the wireless communication system can be enhanced in a variety of scenarios \cite{lywRIS, ganxuRIS1}, especially in the millimeter wave (mmWave) communication scenario.
\par
However, the introduction of RIS has also brought some major challenges \cite{wqqRIS3, robust1, robust2}. Specifically, to maximize the SE of the distributed RIS-based system, the precoding matrix and the phase shifting matrix at the BS and the RIS should be jointly optimized, which was shown as a highly non-convex NP-hard problem \cite{wqqRIS3}, which would lead to significant computational overhead and higher probability of infeasible solutions. Moreover, users usually have high mobility in some typical urban outdoor scenarios, where many machine learning-based algorithms are not applicable any more. In addition, most existing methods assume to have the perfect channel knowledge for beamforming designs, which are usually impractical for RIS-based communication systems \cite{robust1, robust2}. Given the outlined reasons, it is necessary to exploit the new way to implement the beamforming scheme with strong robustness and adaptability for RIS-based communication systems.
\par
There have been some previous works that had paid attention to these problems through various optimization methods \cite{hcwRIS,wqqRIS,guohuayan}. One of the earlier approaches to tackle this joint optimization problem was to use the fractional programming (FP) technique proposed in \cite{hcwRIS}. This approach decoupled the BS precoding and RIS phase optimization problem, however, it might not obtain the globally optimal solution. Similarly, \cite{wqqRIS} divided the problem into two parts: the first part is the standard power minimization problem, while the other one involves the RIS phase optimization. The latter portion was solved utilizing semi-definite relaxation technique, which can achieve high performance, but the computational complexity was high. Another work \cite{guohuayan} presented an alternating optimization (AO) approach for solving the joint design problem in both perfect channel state information (CSI) and imperfect CSI scenarios, where FP technique and successive convex approximation were adopted jointly. This method showed a little adaptability to imperfect CSI, but the performance in dynamic environments is still not guaranteed. 
\par
Recently, the application of deep learning (DL) in wireless communications has gained much attention due to its capability in solving complex and challenging problems \cite{physical_layer, wgan-gp}, since it is naturally adept at extracting valuable features in high-dimensional space with the low complexity. Therefore, a considerable number of works have been dedicated to implementing DL in beamforming across various scenarios \cite{tianlin, universal, xiawenchao, zhangmaojun, unfolding1, unfolding2, unfolding3, unfolding4, unfolding5}. Specifically, \cite{tianlin} proposed a framework to integrate deep neural network (DNN) in beamforming, and transferred the online inference overhead to offline training. However, this approach was limited to single-user cases. To cope with multi-user scenarios, \cite{universal, xiawenchao} presented a model using DNN and custom layers, achieving comparable performance with traditional schemes. In addition, a low-complexity design for beamforming was proposed in \cite{zhangmaojun}, which transformed the multiple-input multiple-output (MIMO) precoding problem into a multiple-input multiple-single (MISO) precoding problem. Furthermore, the deep-unfolding technique has proven to be successful in integrating professional expertise with the feature extraction capabilities of neural networks (NNs) \cite{unfolding1, unfolding2, unfolding3, unfolding4}. It operates by replacing certain parts of the conventional optimization methods with DNNs, resulting in significant overhead reduction and competitive performance compared to traditional methods \cite{unfolding5}. Nevertheless, most of these data-driven approaches are lack of the robustness when dealing with dynamic scenarios, since the black-box nature of DNNs limits the ability to adapt to new data distributions. 
\par
To address the robustness issue, researchers have explored various advanced machine learning techniques \cite{zhufenghao,transfer_learning_survey,MAML1,MAML2,unfolding1, unfolding2, unfolding3, unfolding4, unfolding5,MLAM,mlbf}. Specifically,   \cite{zhufenghao} proposed hybrid learning to tackle the challenge of multiple data sources mixture training, which could make the NNs adaptive in different wireless environments. However, the dynamic nature of wireless channel and inadequate training datasets covering all scenarios limited its practical applications. 
Furthermore, \cite{MAML1} explored the application of transfer learning in downlink beamforming, in which the NNs were trained offline and fine-tuned online, thereby accelerating the online adaptation of NNs. Meanwhile, \cite{MAML2} employed the meta learning to adapt to the change of wireless communication through dual sampling and offline adaptation. However, above mentioned deep learning data-driven techniques usually require the high-quality training and adaptation data, rendering the deployment impractical. To reduce the overhead of pre-training, \cite{MLAM,mlbf} proposed a new optimization framework assisted by long short-term memory networks to replace the weighted minimum mean square error (WMMSE) algorithm \cite{WMMSE}, and simulation results showed that it can successfully improve the overall system SE. But the complexity of this technique remains high, particularly in scenarios involving large antenna arrays.
\par
On the other hand, due to the internal characteristics of the data, redundancy usually exists in the high-dimensional space. As a result, the data that we observe are often projections from a low-dimensional manifold onto a high-dimensional one \cite{ma2011manifold}. Therefore, in many cases, a few lower dimensions data can be a unique representation for high dimension cases. To reduce the data dimensions, manifold learning was proposed in \cite{izenman2012introduction}, which showed that the high dimensional data can be projected to a low dimensional manifold. In light of this, this method has been widely used in wireless-related tasks. Specifically, \cite{location} put forward an innovative compressed sensing method for the localization in wireless sensor networks by using the manifold learning, resulting in remarkable precision at the low price of communication overhead. Moreover, \cite{manifoldMIMO} proposed a two-tier beamforming scheme for massive MIMO communications, which utilized the manifold learning to embed the high-dimensional channels into a low-dimensional subspace while maintaining the potential spatial correlation of those channels. It achieved the near-optimal sum rate and considerably higher energy efficiency than the conventional schemes.
\par
Although some recent works \cite{rethinking, mlbf, wang2023energyefficient} showed that these advanced techniques, i.e., meta learning and manifold learning were also utilized to solve the power allocation problem, they are still in early stages to integrate them into deep learning framework for wireless communications. To the best of our knowledge, there is currently no solution to obtain the optimal joint design of both the BS precoding matrix and the RIS phase shifting matrix, especially in imperfect CSI setups and dynamic scenarios. Attempting to solve this problem in this paper, we propose a pre-training free framework that can achieve the high performance for RIS-aided communication systems with the strong robustness to the diversity of environments as well as the extremely low computational overhead. Our contributions can be summarized as follows.
\begin{itemize}
\item
We propose a gradient-based meta learning method, which works without pre-training and has strong robustness for RIS-aided communication systems. Unlike traditional DL based methods which directly take CSI as input, we take the gradients of the precoding matrix and phase shifting matrix as the input. Moreover, we update the parameters of NNs globally rather than using the traditional, greedy AO methods. It can reduce the risk of falling into local optima and enhance performance over conventional AO methods, while greatly reducing the computational overhead.
\item We propose a manifold learning method to optimize the precoding matrix. Rather than directly optimize the complete precoding matrix at the BS, we only need to optimize it on a low-dimensional manifold. This approach reduces the search space and convergence time, ultimately decreasing overall complexity.
\item Numerical results exhibit that the proposed GMML algorithm outperforms conventional algorithms and existing DL methods in both perfect and imperfect CSI setups. Specifically, compared to traditional schemes, GMML improves the SE by up to 7.31\% and speeds up the convergence up to 23 times faster. Besides, it shows superior performance in adapting to dynamic scenarios. Furthermore, we support reproducible research by providing source code at GitHub. The code for this paper is available online in \cite{sourcecode}.
\end{itemize}
\par
The paper is organized as follows: Section \ref{sec:sys} presents an introduction to the system model and problem formulation. Section \ref{sec:gmml} introduces the GMML framework, followed by the presentation of numerical results in Section \ref{sec:simulation}. Finally, Section \ref{sec:conclusion} concludes the paper.
\par
\textit{Notation}: Fonts $a$, $\mathbf{a}$, and $\mathbf{A}$ denote scalars, vectors, and matrices, respectively. $\mathbf{A}^T$, 
$\mathbf{A}^H$, $\mathbf{A}^{-1}$, and $\|\mathbf{A}\|_2$ represent the transpose, conjugate transpose, inverse, and $L^2$ norm of $\mathbf{A}$, respectively. The $(m,n)$-th entry of $\mathbf{A}$ is denoted by $a_{mn}$, and $|\cdot|$ denotes the modulus. $\mathcal{O}$ represents the asymptotic time complexity. Finally, we can represent the diagonal matrix and trace of matrix $\mathbf{A}$ using the notations $\mathrm{diag}(\mathbf{a})$ and $\mathrm{Tr}(\mathbf{A})$, respectively.
\par



\vspace{2mm}
\section{System Model and Problem Formulation}\label{sec:sys}
In this section, we present the system model, followed by the formulation of the SE maximization problem. 

\begin{figure}[t]\vspace{-0mm}
	\begin{center}		\centerline{\includegraphics[width=0.45\textwidth]{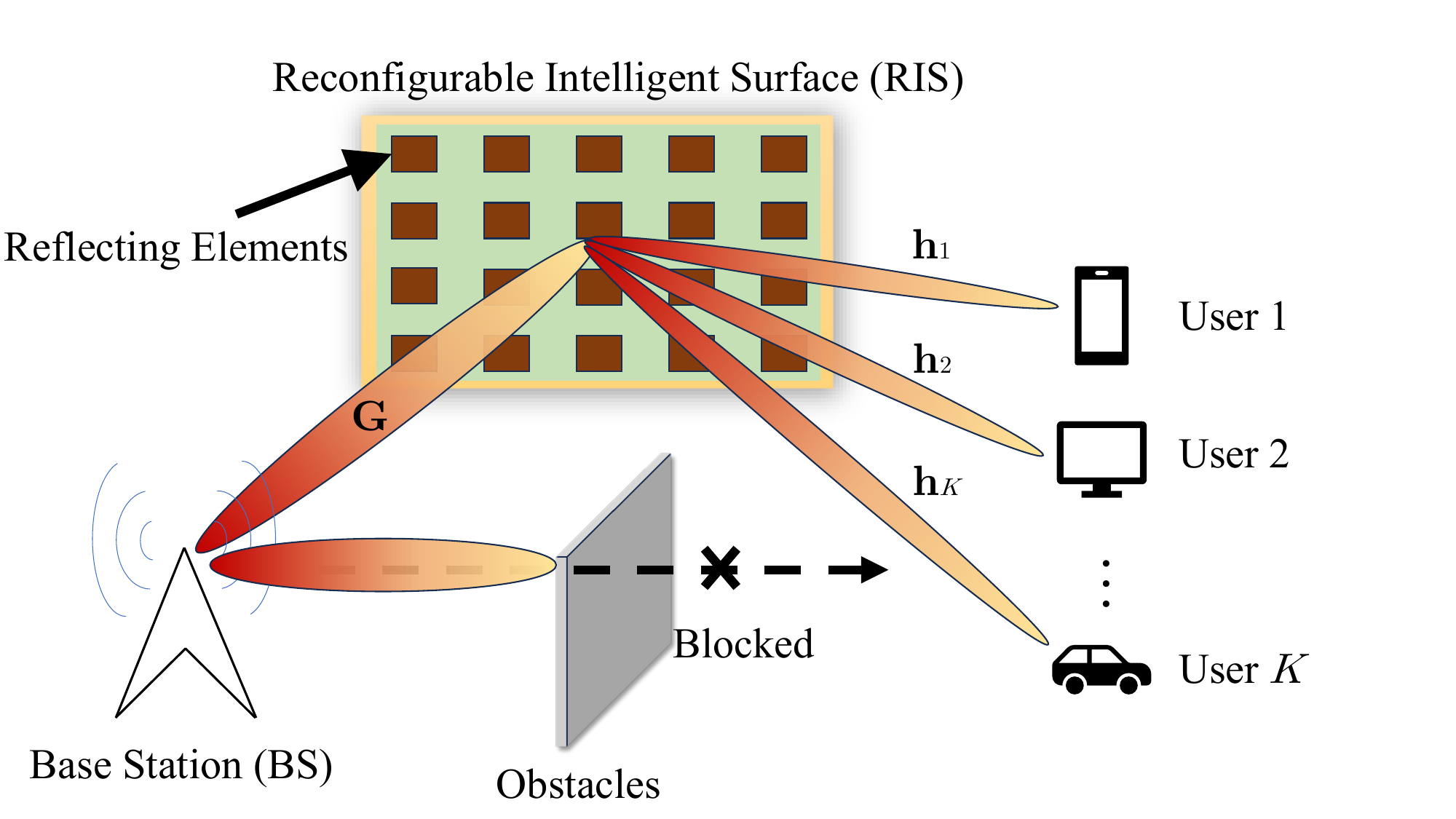}}  \vspace{-0mm}
	    \captionsetup{font=footnotesize, name={Fig.}, labelsep=period}  
		\caption{\, RIS-aided MU-MISO beamforming system.}
		\label{fig:system} \vspace{-10mm}
	\end{center}
\end{figure}

As shown in Fig. \ref{fig:system}, the mmWave MISO communication system consists of one $M$-antenna BS, one RIS with $N$ reflective elements, and $K$ single-antenna users. The data stream of the $k$-th user is marked as $s_k$ and satisfies  $E\left\{| {s}_k |^2 \right\}=1$. Due to the heavy attenuation of the mmWave, the direct links between the BS and the users are negligible. Therefore, the users can only communicate with the BS via the virtual links provided by the RIS. Furthermore, we assume that all the channels, including the channel $\mathbf{G} \in \mathbb{C}^{N \times M}$ between the BS and the RIS, and the channel $\mathbf{H} \in \mathbb C^{K\times N}$ between the RIS and users, are known. Meanwhile, in the imperfect CSI setup, we assume that the channels are known but with estimation error. The signal received by the $k$-th user is denoted as $y_k$ and can be written as
\begin{equation}\label{kth User Received}
	y_k=\mathbf{h}^H_k\mathbf{\Theta G w}_k s_k+\sum_{i\neq k}^K \mathbf{h}^H_k \mathbf{\Theta G w}_i s_i + n_k.
\end{equation}
Here the RIS phase shifting matrix is defined as $\mathbf \Theta=\text{diag}[e^{j\theta_1}, e^{j\theta_2}, \cdots, e^{j\theta_N}]$, where $\theta_n $ represents the phase shift induced by the $n$-th RIS element. The $k$-th column of the precoding matrix $\mathbf W\in \mathbb{C}^{M\times K}$ at the BS is defined as $\mathbf{w}_k$, and $\mathbf{h}_k$ represents the transpose of the $k$-th row of $\mathbf{H}$. The noise $n_k$ for the $k$-th user is additive circular complex white Gaussian noise with zero mean and variance $\sigma^2$. It is noted that our method can also be extended to scenarios with simultaneously transmitting and reflecting RIS aided wireless communications \cite{STAR-RIS1, STAR-RIS2, STAR-RIS3} with similar optimization methods.
\par 
Practically, the total power of all antennas in a BS is limited. Therefore, a constraint on $\mathbf{W}$ is provided as
\begin{equation}\label{Power Constraint}
	\mathrm{Tr}(\mathbf{W}^ H\mathbf{W})\leq P,
\end{equation}
where $P$ is the total maximum transmit power of the BS. In \eqref{kth User Received}, the term $\mathbf{h}^H_k\mathbf{\Theta G w}_k s_k$ is the desired signal at the $k$-th user while $\sum_{i\neq k}^K \mathbf{h}^H_k \mathbf{\Theta G w}_i s_i$ is treated
as the interference between the $k$-th user and the other users. Since $s_k$ is unit-powered, the signal to interference plus noise ratio (SINR) for the $k$-th user can be expressed as:
\begin{equation}\label{snr single user}
	\gamma_k = \frac{|\mathbf{h}_k^H\mathbf{\Theta G w}_k|^2}{\sigma^2 + \sum_{j \neq k}^K|\mathbf{h}_k^H\mathbf{\Theta G w}_j|^2}.
\end{equation}
To evaluate the system performance, the SE serves as a metric, which can be quantified as
\begin{equation}\label{spectrum efficiency}
	R(\mathbf{W,\Theta;H,G})=\sum_{k=1}^K \omega_k \log_2(1+\gamma_k).
\end{equation}
Here $\omega_k$ refers to the weight of the $k$-th user. The aim of this paper is to maximize the target function as described in \eqref{spectrum efficiency} by optimizing $\mathbf \Theta$ and $\mathbf W$ jointly. Therefore, the optimization problem could be expressed as
\begin{equation}\label{optimization problem}
    \begin{split}
        \mathop{\rm {max}}\limits_{\substack{\mathbf{W}\in \mathcal{W}\; \\ \mathbf{\Theta}\in\mathcal{O}}} & R(\mathbf{W,\Theta;H,G}),\\
	     \mathrm{s.t.\ \ }&\mathrm{Tr}(\mathbf{W}^{H}\mathbf{W})\leq P,\\
	    &\mathbf{\Theta}=\mathrm{diag}[e^{j\theta_1},e^{j\theta_2},\cdots,e^{j\theta_N}],\\
	    &|\theta_j|\in [0, 2\pi), j=1,2,3,\cdots,N.
    \end{split}
\end{equation}
where $\mathcal{W}$ and $\mathcal{O}$ are the feasible regions of $\mathbf{W}$ and $\mathbf{\Theta}$, respectively.
To assess the channel estimation error (CEE), we refer to the true channel as $\mathbf{h}$ and the estimated channel as $\hat{\mathbf{h}}$. Then the CEE can be measured in decibels (dB) as follows
\begin{equation}\label{CEE}
    \mathrm{CEE} = 10\log_{10}\left(\frac{\mathbb{E}[\| \mathbf{h} -  \hat{\mathbf{h}} \|_2^2]}{\mathbb{E}[\| \mathbf{h} \|_2^2]}\right).
\end{equation}
CEE assesses the accuracy of channel estimation. The smaller the value of CEE, the more precise the channel estimation is.
The computational complexity and accuracy of channel estimation directly affect the overall system efficiency, therefore, it is important to acquire accurate CSI with low overhead. However, acquiring accurate CSI is a challenging task in RIS-aided communications, as it involves estimating $\mathbf{H}$ and $\mathbf{G}$ simultaneously in a dynamic environment, and the dimensions of $\mathbf{H}$ and $\mathbf{G}$ grow linearly with $N$. Currently, there are two primary methods to estimate the CSI as follows:
\subsubsection{Compressed Sensing Method} In \cite{compressed}, a compressed sensing method for channel estimation was proposed, where a data-driven and compressive sensing based approach was developed with low training overhead. The technique exploited the common sparse properties between the different sub-carriers and the double-structured sparse properties of the cascaded channel matrices in the angular space.
\subsubsection{Tensor-based Method} In \cite{hcwCE}, two iterative algorithms were proposed for RIS-aided communications. They used parallel factor decomposition to unfold the cascaded channel into three different modes. And the proposed tensor-based methods have been proven effective and robust in various scenarios. 

\vspace{0mm}

\section{GMML Framework}\label{sec:gmml}
In this section, we describe the fundamentals of the GMML framework. The first is the manifold learning technique, while the second is the model-driven gradient-based meta learning method. 
In the following subsections, we provide details on each of the two methods.
\subsection{Manifold Learning}\label{sec:manifold}
In the traditional optimization algorithms (e.g., AO), high-dimensional matrix inversion is required in each iteration to optimize the precoding matrix $\mathbf{W}$, which would result in a cubic computational complexity with respect to $M$. When the BS is equipped with a large number of antennas, the overhead of matrix inversion would become excessively high. On the other hand, in the context of DL, NNs have been introduced to replace the matrix inversion, but the dimension of the search space for the precoding matrix $\mathbf{W}$ remains high. As a result, training these networks is a demanding and expensive task with a high risk of over-fitting.
\par
To tackle these challenges, we leverage the manifold learning method to simplify \eqref{optimization problem} into a lower-dimensional problem. First, we introduce a property indicating that full power is required to maximize SE. Denote the concatenated channel of the $k$-th users as $\mathbf{h}_{c,k}=\mathbf{h}^H_k \mathbf{\Theta G} \in \mathbb{C}^{1 \times M}$, and the cascaded channel of all users as $\mathbf{H}_c=[\mathbf{h}_{c,1}^H, \mathbf{h}_{c,2}^H,\cdots,\mathbf{h}_{c, K}^H]^H \in \mathbb{C}^{K \times M}$. Any $\mathbf{W}$ that satisfies $|\mathbf{h}_k^H\mathbf{\Theta G w}_k|^2=0, \forall k$ is defined as a trivial stationary point. With the definition of a trivial stationary point, the following proposition holds for the problem \eqref{optimization problem}.
\par
\textit{Proposition 1:} Any nontrivial stationary point \{$\mathbf{w}^*_k$\} of problem \eqref{optimization problem} must conform the power constraint with equality.
\par
\textit{Proof}: See Appendix \ref{appendix1}.
\par
This property of power constraint equality eliminates the need for optimal power search through bisection method in traditional algorithms. It indicates that we only need to constraint the transmit power to the maximum achievable value after optimizing the precoding matrix $\mathbf{W}$ in NNs. With \textit{Proposition 1}, we can derive another important property.
\par
\textit{Proposition 2:} Any nontrivial stationary point $\{ \mathbf{w}^*_k \}$ of problem \eqref{optimization problem} must lie in the range space of $\mathbf{H}_c^H$, satisfying $\mathbf{w}^*_k = \mathbf{H}_c^H\mathbf{X}_k$ with unique $\mathbf{X}_k \in \mathbb{C}^{K \times 1}$.
\par
\textit{Proof}: See Appendix \ref{appendix2}.
\par
This proposition implies that the search space of the original precoding matrix $\mathbf{W} \in \mathbb{C}^{M \times K}$ can be reduced to a much smaller compressed precoding matrix $\mathbf{X}=[\mathbf{X}_1,\mathbf{X}_2,\cdots,\mathbf{X}_K] \in \mathbb{C}^{K \times K}$. This is because that $M$ is usually much larger than $K$ when large antenna arrays are deployed at the BS. Consequently, we can present the transformed optimization problem as follows
\begin{equation}\label{transformed optimization problem}
    \begin{split}
        \mathop{\rm {max}}\limits_{\substack{\mathbf{W}\in \mathcal{W}\; \\ \mathbf{\Theta}\in\mathcal{O}}} & R(\mathbf{H}_c^H \mathbf{X,\Theta;H,G}),\\
	     \mathrm{s.t.\ \ }&\mathrm{Tr}(\mathbf{X}^H \mathbf{H}_c\mathbf{H}_c^H\mathbf{X}) = P,\\
	    &\mathbf{\Theta}=\mathrm{diag}[e^{j\theta_1},e^{j\theta_2},\cdots,e^{j\theta_N}],\\
	    &|\theta_j|\in [0, 2\pi),j=1,2,3,\cdots,N.
    \end{split}
\end{equation}
Note that the optimal solution of \eqref{optimization problem} is also a stationary point, according to \textit{Proposition 1} and \textit{Proposition 2}. This, in turn, suggests that the optimal solution can be attained by optimizing $\mathbf{X}$ on a low-dimensional manifold. The proposed GMML framework exploits this property by optimizing the compressed precoding matrix through NNs instead of directly optimizing the original precoding matrix. As a result, the complexity of searching for the optimal parameters is reduced, and the energy efficiency is enhanced.

\subsection{Gradient Based Meta Learning}
In this subsection, we would introduce the gradient-based optimization method and the meta learning architecture, as well as the details of the NNs.
\subsubsection{Gradient As Input}
Traditional DL-based methods typically input the matrices $\mathbf{H}$ and $\mathbf{G}$ directly into the NNs, which then output the target matrices $\mathbf{W}^*$ and $\mathbf{\Theta}^*$. However, the optimization process within the black-box DNN remains not fully understood. Inspired by the gradient descent method, we feed the gradients $\nabla_\mathbf{X} R$ and $\nabla_\mathbf{\Theta} R$ into the NNs and the outputs $\Delta \mathbf{X}$ and $\Delta \mathbf{\Theta}$ would be added to the initialized or updated $\mathbf{X}$ and $\mathbf{\Theta}$. Compared with traditional DL techniques, this approach offers improved interpretability. Besides, it extracts higher order information of the matrices $\mathbf{X}$ and $\mathbf{\Theta}$, which can result in better optimization to both variables. Moreover, we utilize the existing automatic differentiation mechanism of PyTorch to calculate the gradients in practice. This mechanism efficiently aids in obtaining the gradients.
\begin{figure}\vspace{-0mm}
	\begin{center}
		\centerline{\includegraphics[width=0.5\textwidth]{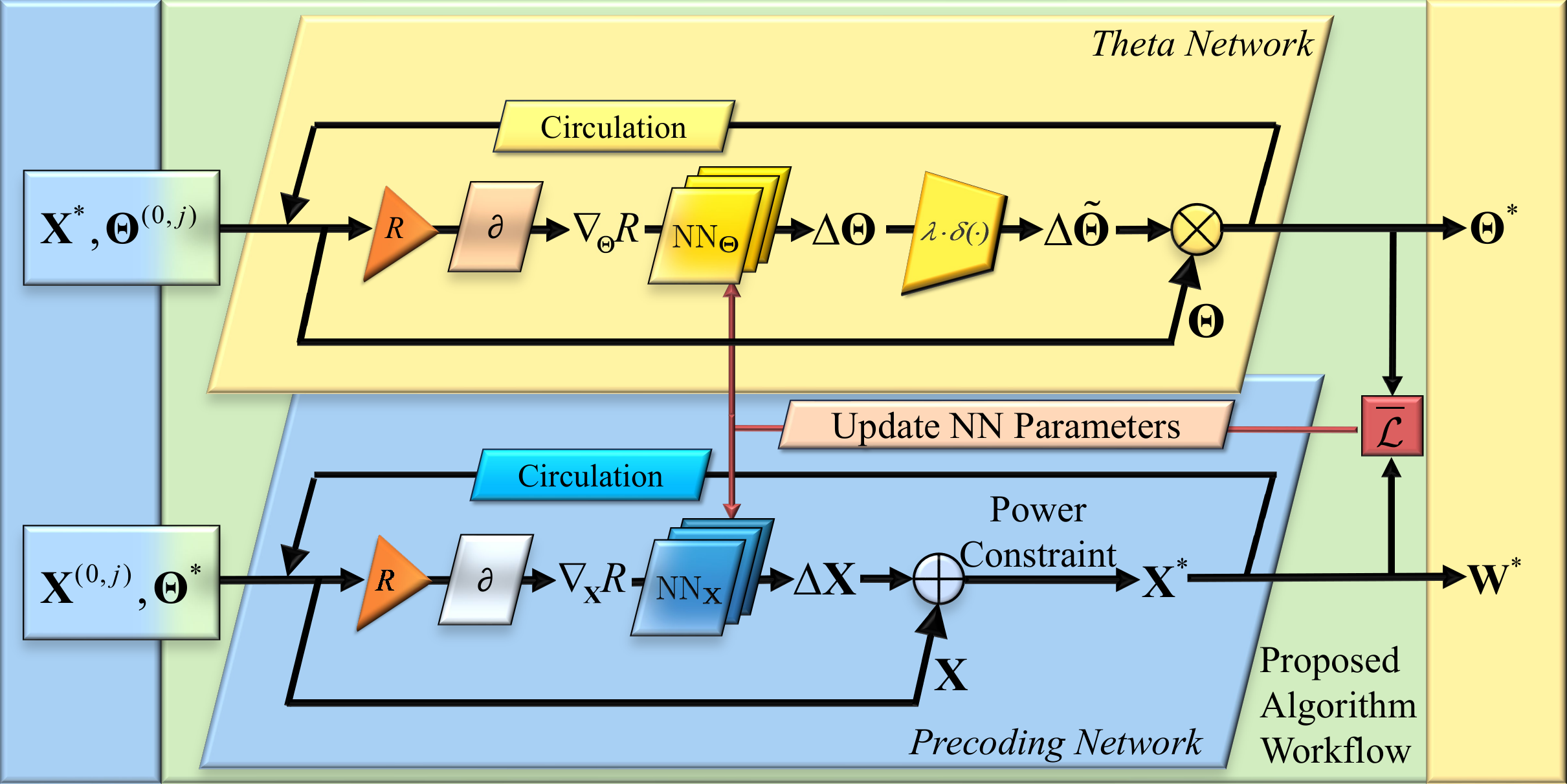}}  \vspace{-0mm}
		\captionsetup{font=footnotesize, name={Fig.}, labelsep=period} 
		\caption{\, GMML architecture.}
		\label{fig:framework} \vspace{-10mm}
	\end{center}
\end{figure}
\subsubsection{Meta Learning Architecture}
Traditional data-driven meta learning methods typically necessitate substantial offline pre-training and further online adaptation refinement, such as model-agnostic meta learning \cite{finn2017model, MAML1}. However, these methods are sensitive to CEE and variation in data distribution. Additionally, performing large-scale pre-training and frequent adaptation consumes significant amount of energy, rendering this approach unsuitable for latency-critical and dynamic scenarios. To tackle these issues, we propose a pre-training free, model-driven meta-learning framework that exhibits strong robustness, whose primary optimization target is the search trajectory rather than individual variables. The algorithm features a three-layer nested cyclic optimization structure, which comprises epoch iterations, outer iterations, and inner iterations. The details are as follows.
\begin{itemize}
    \item \emph{Inner iteration:} This iteration is responsible for optimizing the target variables cyclically. As shown in Fig. \ref{fig:framework}, there are two sub-networks responsible for optimizing $\mathbf{W}$ and $\mathbf{\Theta}$ separately in the inner iteration, called \textit{Precoding Network (PN)} and \textit{Theta Network (TN)}. In each sub-network, we update $\mathbf{\Theta}$ and $\mathbf{X}$ sequentially in each inner iteration, and the the target optimization variable is inherited from initial while the other variable is inherited from the other sub-network. The update process in the $j$-th outer iteration can be formulated as 
\begin{align}
    \mathbf{X}^* & = {PN}(\mathbf{X}^{(0,j)}, \mathbf{\Theta}^*), \label{update X}\\
    \mathbf{\Theta}^* & = {TN}(\mathbf{X}^*, \mathbf{\Theta}^{(0,j)}),
\end{align}
where $\mathbf{X}^{(i,j)}$ and $\mathbf{\Theta}^{(i,j)}$ represent $\mathbf{X}$ and $\mathbf{\Theta}$ in the $i$-th inner iteration of the $j$-th outer iteration. 
\item \emph{Outer iteration}: This iteration is responsible for accumulating the loss, and there are $N_i$ inner iterations in each outer iteration. 
To work in a self-supervised learning manner, the loss function in $j$-th outer iteration can be expressed as the negative value of SE
\begin{equation}\label{Loss Function}
    \mathcal{L}^j= - R(\mathbf{W}^*,\mathbf{\Theta}^{*};\mathbf{H}, \mathbf{G}).
\end{equation}
\item \emph{Epoch iteration}: This iteration is in charge of updating the parameters of NNs, and there are $N_o$ outer iterations in each epoch iteration. After completing $N_o$ outer iterations, the losses are summed and averaged as
\begin{equation}\label{Average Loss Function}
    \overline{\mathcal{L}}= \frac{1}{N_o}\sum_{j = 1}^{N_o}\mathcal{L}^j.
\end{equation}
Then backward propagation is conducted and the Adam optimizer is used to update the NNs embed in both sub-networks, as depicted below
\begin{align}
    \mathbf{\theta_X}^* & =\mathbf{\theta_X} + \alpha_{\mathbf{X}} \cdot \mathrm{Adam}(\nabla_{\mathbf{\theta_X}}\overline{\mathcal{L}}, \mathbf{\theta_{X}}), \label{update_X_NN_parameter} \\
    \mathbf{\theta_\Theta}^* & =\mathbf{\theta_\Theta} + \alpha_{\mathbf{\Theta}} \cdot \mathrm{Adam}(\nabla_{\mathbf{\theta_\Theta}}\overline{\mathcal{L}},\mathbf{\theta_\Theta} ), \label{update_theta_NN_parameter}
\end{align}
where $\alpha_{\mathbf{X}}$ and $\alpha_{\mathbf{\Theta}}$ are the learning rates of the two networks, respectively. There are $N_e$ epoch iterations in the whole optimization process, and the update interval of \eqref{update_theta_NN_parameter} is set to a constant $n_0$ to balance the alternative optimization. Therefore, there is one update to the parameters of the \textit{PN} and the \textit{TN} in one and $n_0$ epoch iterations, respectively. It globally controls the optimization direction of $\Delta \mathbf{X}$ and $\Delta \mathbf{\Theta}$, thus being less greedy and more efficient than the AO method.
\end{itemize}
The meta-learning architecture learns to learn through the following specific process. Consider equation \eqref{update X} as an illustration: the objective of the $PN$ is to optimize the precoding matrix. At the outset of each outer iteration, the original randomly initialized precoding matrix \(\mathbf{X}^{(0,j)}\) is fed into the $PN$. In this network, the precoding matrix does not inherit from the previous outer iteration; instead, only the phase shifting matrix inherits the optimized matrix from prior iterations. Conversely, for the $TN$, the phase shifting matrix does not inherit from the previous outer iteration; rather, only the precoding matrix inherits the optimized matrix from earlier iterations. This ensures that one variable is optimized from scratch within each respective network in the outer iteration. The inner iterations within one outer iteration act as optimization steps, forming an optimized trajectory. In our algorithm, this trajectory is continually updated and optimized, enabling the learning of an effective strategy.

\subsubsection{Precoding Network}
As illustrated in \ref{sec:manifold}, the precoding matrix $\mathbf{w}_k$ is in the range space of the cascaded channel $\mathbf{H}_c$, we only need to optimize the compressed precoding matrix $\mathbf{X}$ instead of the original $\mathbf{W}$. Denote SE in the $i$-th inner iteration and $j$-th outer iteration as $R_{\mathbf{X}}^{(i, j)}$, which can be expressed as follows 
\begin{equation}\label{updateX}
	R_{\mathbf{X}}^{(i, j)}=\sum_{k=1}^K \omega_k \log_2(1 + \frac{|\mathbf{h}_k^H \hat{\mathbf{\Theta}} \mathbf{G} \mathbf{H}_c^H\mathbf{X}_k^{(i, j)}|^2}{\sigma^2 + \sum_{j \neq k}^K|\mathbf{h}_k^H \hat{\mathbf{\Theta}} \mathbf{G} \mathbf{H}_c^H\mathbf{X}_j^{(i, j)}|^2}),
\end{equation}
where $\hat{\mathbf{\Theta}}$ is either the initialized or updated phase shifting matrix. As shown in Fig. \ref{fig:framework}, in the workflow of the \textit{PN}, SE is first computed and the gradient of $\mathbf{X}^{(i, j)}$ with respect to SE is fed into the light-weighted NNs, then the output $\Delta \mathbf{X}^{(i, j)}$ is added to $\mathbf{X}^{(i, j)}$ and regulated to satisfy power constraint, leading to the updated $\mathbf{X}^{(i+1, j)}$
\begin{align}
    \mathbf{X}^* & = \mathbf{X}^{(i, j)} + \Delta \mathbf{X}^{(i, j)}, \\
    \mathbf{X}^{(i+1, j)} & = \sqrt{\frac{P}{\mathrm{Tr}(\mathbf{H}_c^H\mathbf{X}^*(\mathbf{H}_c^H\mathbf{X}^*)^H)}}\mathbf{X}^*. \label{antenna_power_constraint}
\end{align}
Afterwards, the precoding matrix is recovered by multiplying cascaded channel to the compressed precoding matrix
\begin{equation}\label{precoding_matrix_recovery}
    \mathbf{W}^{(i+1, j)} = \mathbf{H}_c^H\mathbf{X}^{(i+1, j)}.
\end{equation}
The procedure described above is iterated $N_i$ times in an inner iteration, which is similar to $N_i$ steps in traditional gradient descend method. However, thanks to the feature extraction ability of NNs and the information fusion ability of meta learning, the gradient information flow can be shared and modified across different networks. This allows for the coupled variables in the highly non-convex optimization problem to be handled in a discrete manner.

\begin{algorithm}[t]
\caption{GMML Workflow }
\label{alg:meta}
\begin{algorithmic}[1]
\Procedure{GMML}{$\mathbf{H}$, $\mathbf{G}$}
    \State Randomly Initialize $\mathbf{\theta}_{\mathbf{X}},\mathbf{\theta}_{\mathbf{\Theta}},\mathbf{X}^{(0,1)}, \mathbf{\Theta}^{(0,1)}$.
    \State Normalize $\mathbf{X}^{(0,1)}$ as \eqref{antenna_power_constraint}.
    \State Initialize $\mathbf{W}^{(0,1)}$ as \eqref{precoding_matrix_recovery} with $\mathbf{X}^{(0,1)}$.
    \State Initialize the maximum SE recorder as $\mathrm{MAX}=0$.
    \For{$k\leftarrow 1,2,\cdots,N_e$}
    \State $\overline{\mathcal{L}}=0$;
    \For{$j\leftarrow 1,2,\cdots,N_o$}
    \State $\mathbf{\Theta}^{(0,j)}=\mathbf{\Theta}^{(0,1)}$;
    \State $\mathbf{X}^{(0,j)}=\mathbf{X}^{(0,1)}$;
    \State $\mathbf{W}^{(0,j)}=\mathbf{W}^{(0,1)}$.
    \For{$i\leftarrow 1,2,\cdots,N_i$}
        \State $R^{(i-1, j)}_{\mathbf{\Theta}} = R(\mathbf{\mathbf{W}^*, \Theta}^{(i-1, j)}; \mathbf{H, G})$;
        \State $\Delta \mathbf{\Theta}^{(i-1, j)} = \mathrm{NN}_{\mathbf{\Theta}}(\nabla_{\mathbf{\Theta}} R^{(i-1, j)})$ ;
        \State $\Delta \widetilde{\mathbf{\Theta}}^{(i-1, j)} = \lambda \cdot \delta( \Delta \mathbf{\Theta}^{(i-1, j)})$ ;
        \State $\mathbf{\Theta}^{(i, j)}\leftarrow \mathbf{\Theta}^{(i-1, j)} \cdot \Delta \widetilde{\mathbf{\Theta}}^{(i-1, j)}$.
    \EndFor
    \State $\mathbf{\Theta}^*  = \mathbf{\Theta}^{(N_i,j)}$;
    \For{$i\leftarrow 1,2,\cdots,N_i$}
        \State $R^{(i-1, j)}_{\mathbf{X}} = R(\mathbf{W}^{(i-1, j)}, \mathbf{\Theta}^{*}; \mathbf{H, G})$;
        \State $\Delta \mathbf{X}^{(i-1, j)} = \mathrm{NN}_{\mathbf{X}}(\nabla_{\mathbf{X}} R^{(i-1, j)}_{\mathbf{X}})$ ;
        \State $\mathbf{X}^{(i, j)}\leftarrow \mathbf{X}^{(i-1, j)}+\Delta \mathbf{X}^{(i-1, j)}$;
        \State Normalize $\mathbf{X}^{(i, j)}$ as \eqref{antenna_power_constraint};
        \State $\mathbf{W}^{(i, j)} = \mathbf{H}_c \mathbf{X}^{(i, j)}$.
    \EndFor
    \State $\mathbf{\mathbf{X}}^*  = \mathbf{\mathbf{X}}^{(N_i,j)}$;
    \State $\mathbf{\mathbf{W}}^*  = \mathbf{\mathbf{W}}^{(N_i,j)}$;
    \State $\mathcal{L}^j= - R(\mathbf{W}^*,\mathbf{\Theta}^*;\mathbf{H}, \mathbf{G})$;
    \State $\overline{\mathcal{L}} = \overline{\mathcal{L}} + \mathcal{L}^j$.
    \If{$-\mathcal{L}^j > \mathrm{MAX}$} 
        \State $\mathrm{MAX} = -\mathcal{L}^j$;
        \State $\mathbf{W}_{opt} = \mathbf{W}^*$;
        \State$\mathbf{\Theta}_{opt} = \mathbf{\Theta}^* $.
    \EndIf
    \EndFor
    \State $\overline{\mathcal{L}} = \frac{1}{N_o}\overline{\mathcal{L}}$;
    \State Update $\mathbf{\theta}_{\mathbf{X}}$ as \eqref{update_X_NN_parameter}.
     \If{$N_e \mid n_0$} 
     \State Update $\mathbf{\theta}_{\mathbf{\Theta}}$ as \eqref{update_theta_NN_parameter}.
     \EndIf
    \EndFor
    \State \Return $\mathbf{W}_{opt}$, $\mathbf{\Theta}_{opt}$.
\EndProcedure
\end{algorithmic}
\end{algorithm}

\subsubsection{Theta Network}
Similar to \eqref{updateX}, we denote SE in the $i$-th inner iteration and $j$-th outer iteration as $R_{\mathbf{\Theta}}^{(i, j)}$, which can be expressed as follows
\begin{equation}\label{updateTheta}
	R_{\mathbf{\Theta}}^{(i, j)}=\sum_{k=1}^K \omega_k \log_2(1 + \frac{|\mathbf{h}_k^H \mathbf{\Theta}^{(i, j)} \mathbf{G} \mathbf{H}_c^H \hat{\mathbf{X}}_k|^2}{\sigma^2 + \sum_{j \neq k}^K|\mathbf{h}_k^H \mathbf{\Theta}^{(i, j)} \mathbf{G} \mathbf{H}_c^H\hat{\mathbf{X}}_j|^2}),
\end{equation}
where $\hat{\mathbf{X}}$ is either the initialized or the updated precoding matrix. Although \eqref{updateTheta} shares a similar formulation with \eqref{updateX}, the behavior of the target optimization variable is quite different, which presents a major challenge. Specifically, if the RIS phase shifting matrix is updated directly with  $\mathbf{\theta}^{(i+1, j)}=\mathbf{\theta}^{(i, j)} + \Delta \mathbf{\theta}$,
the change in $\mathbf{\Theta}^{(i+1, j)}$ with $\Delta \mathbf{\theta}$ may not be monotonic due to the periodicity of trigonometric functions. Consequently, the output of the NN would exceed its intended range, leading to fluctuations in the SE, and causing difficulty in achieving convergence. This poses a challenge in determining the optimal point, therefore, it is crucial to address this issue in the design of \textit{Theta Network} to improve stability during iterations. It is a well-established fact that the period of a trigonometric function is $2\pi$. In light of this, we have designed a customized regulator to ensure that $\Delta\mathbf{\theta}$ is constrained within the range of $0$ to $2\pi$, expressed as $\Delta \mathbf{\widetilde{\theta}}=\lambda \cdot \delta( \Delta \mathbf{\theta})$,
where $\lambda$ acts as an amplification operator and $\delta(\cdot)$ represents the sigmoid function, and this operation is carried out on the diagonal components of $\mathbf{\Theta}$. This design ensures that the $\mathbf{\Theta}$ is within the limited range in an inner iteration, thus the updated phase shifting matrix can be expressed as
\begin{equation}\label{differential constraint}
\begin{split}
    \mathbf{\Theta}^{(i+1, j)}  & = \mathrm{diag}[e^{j(\mathbf{\theta}^{(i, j)} + \Delta \mathbf{\widetilde{\theta}})}] \\
    & = \mathrm{diag}[e^{j \mathbf{\theta}^{(i, j)}}] \cdot \mathrm{diag}[e^{j(\lambda \cdot \delta( \Delta \mathbf{\theta}))}] \\
    & = \mathbf{\Theta}^{i} \cdot \Delta \mathbf{\widetilde{\Theta}}.
\end{split}
\end{equation}
The number of neurons in each layer of both NNs are listed in Table \ref{NN_Detail}, and the algorithm is detailed in Algorithm \ref{alg:meta}.

\subsection{Discussion}
Thanks to the manifold learning, the dimension of search space for the precoding matrix at the BS decreases from $MK$ to $K^2$. In addition, the gradient-based meta learning approach avoids the matrix inversion, which further decreases the complexity. The complexity of GMML has been proven to be $\mathcal{O}(N_e N_o N_i K^2 M N)$, with further details available in Appendix \ref{appendix3}. Besides, the complexity of AO has shown to be $\mathcal{O}(L_3(L_1 K M^3 + L_2 K^2 N^2))$ \cite{guohuayan}. Compared with AO, GMML simplifies the complexity of $M$ from cubic to linear and the complexity of $N$ and from quadratic to linear in a single iteration, indicating that it is insensitive to the number of antennas or RIS elements.

\vspace{0mm}
\section{Numerical Results}\label{sec:simulation}
\begin{figure}[t]\vspace{4mm}
	\begin{center}
		\centerline{\includegraphics[width=0.45\textwidth]{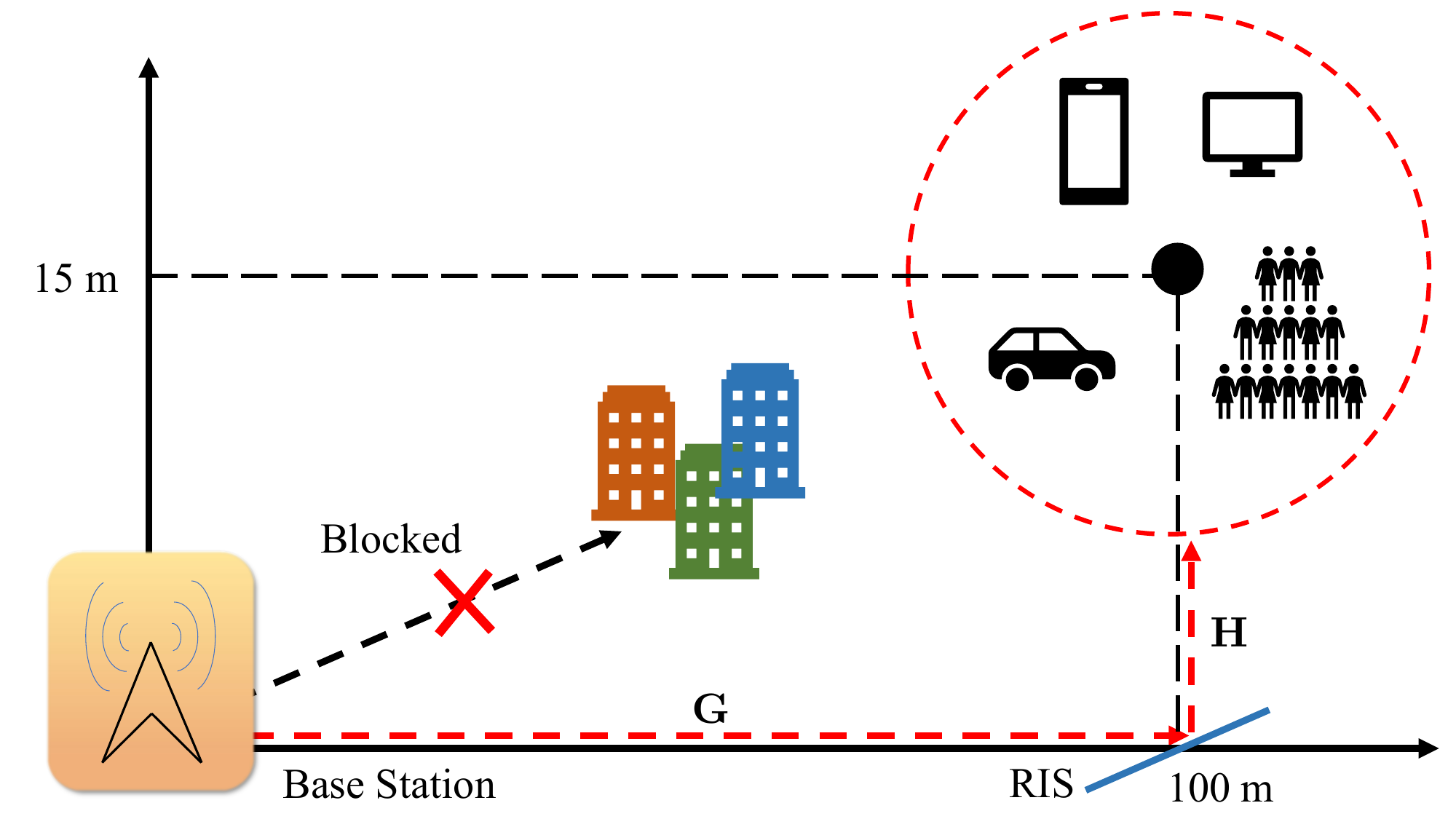}}  \vspace{-0mm}
		\captionsetup{font=footnotesize, name={Fig.}, labelsep=period} 
		\caption{\, The simulation scenario for MISO communication aided by RIS includes $K$ users, a $M$-antenna AP, and an $N$-element RIS.}
		\label{fig:simulation_scene} \vspace{-5mm}
	\end{center}
\end{figure}
In this section, we demonstrate the performance of the proposed method through simulation results. The simulation scenario for RIS-aided MISO communication is illustrated in Fig. \ref{fig:simulation_scene}. The base station is located at (0 m, 0 m) and communicates with 4 users who are randomly positioned within a circle centered at (100 m, 15 m) with a radius of 5 meters. Due to obstacles that block the direct links, the users can only communicate with the BS through the passive reflection of RIS, which is located in (100 m, 0 m). More system parameters can be found in Table \ref{tab:simulation parameters}, in which the path loss is established based on the 3GPP propagation environment standard \cite{3gpp.36.814}, and $\lambda$ stands for the wavelength.
\par
We assume that the channel between the RIS and users and the channel between the RIS and the BS both follow the Rician fading, which are modeled as
\begin{align}
    \mathbf{h}_{k} & =L_{h_k}^{LoS}\sqrt{\frac{\kappa_{h}}{1+\kappa_{h}}}\mathbf{h}_{k}^{LoS}
    +L_{h_k}^{NLoS}\sqrt{\frac{1}{1+\kappa_{h}}}\mathbf{h}_{k}^{NLoS},\label{channel_h}\\
    \mathbf{G} & =L_G^{LoS}\sqrt{\frac{\kappa_G}{1+\kappa_G}}\mathbf{G}^{LoS}
    +L_G^{NLoS}\sqrt{\frac{1}{1+\kappa_G}}\mathbf{G}^{NLoS},\label{channel_G}
\end{align}
where $L_{h_k}^{LoS}$, $L_{h_k}^{NLoS}$, $L_{G}^{LoS}$ and $L_G^{LoS}$ represent the path losses of channels, respectively. And we set the Rician factors $\kappa_{h}$ and $\kappa_G$ both to 10. The line of sight (LoS) paths are represented by $\mathbf{h}_{k}^{LoS}$ and $\mathbf{G}^{LoS}$, while the non line of sight (NLoS) paths are represented by $\mathbf{h}_{k}^{NLoS}$ and $\mathbf{G}^{NLoS}$. For the imperfect CSI setup, we assume that the estimation error $\mathbf{z} = \mathbf{h} - \hat{\mathbf{h}}$ follows zero mean additive circular complex white Gaussian distribution.
\par
In the simulations, we compare our results with several other algorithms on $N_s$ independent channel samples. The system parameters $N_s, N_e, N_o, N_i, \alpha_{\mathbf{W}}, \mathbf{\alpha_{\Theta}}, \lambda$ and $n_0$ are set as $100, 500, 1, 1, 1\times 10^{-3}, 1.5\times10^{-3}, 2\pi$ and $5$, respectively. We run the simulations on a computer equipped with an EPYC 75F3 CPU and a RTX 3090 GPU using PyTorch 2.0.1 and Python 3.9. To serve as baselines, other algorithms are listed below, and the computational complexity is summarized in Table \ref{complexity}.
\begin{itemize}
    \item \textbf{Baseline 1} (Random Phase): $\mathbf{\Theta}$ is randomly initialized and \eqref{optimization problem} is solved by the WMMSE algorithm.
    \item \textbf{Baseline 2} (GML): Gradient-based meta learning (GML) is a simplified version of GMML which removes the manifold learning technique but leaves the rest of GMML unchanged.
    \item \textbf{Baseline 3} (ML): Meta learning (ML) is a simplified version of GMML which removes the manifold learning technique and gradient input mechanism.
    \item \textbf{Baseline 4} (DNN): The deep learning model in \cite{xu2021robust} is adopted, which utilizes a black-box DNN to optimize  $\mathbf{\Theta}$ and $\mathbf{W}$ simultaneously.
    \item \textbf{Baseline 5} (AO): The traditional optimization method in \cite{guohuayan} is adopted, which optimizes $\mathbf{\Theta}$ and $\mathbf{W}$ with the manifold optimization method and WMMSE method alternatively and greedily.
\end{itemize}

\begin{table}[t] \vspace{-0mm}
    \centering
    \caption{Number of neurons in the NNs}\vspace{0mm}
    \label{NN_Detail}
    \begin{tabular}{c c c c}
     \toprule
     No.&Layer Name &\emph{Precoding-Network}&\emph{Theta-Network}  \\
     \midrule
     1 & Input Layer& $2\times K$ & $N$\\
     2 & Linear Layer 1 & $200$   & $200$\\
     3 & ReLU Layer & 200 & 200 \\
     4 & Output Layer & $ 2\times K$ & $N$\\
     5 & Differential Regulator & / & $N$\\ [1.0ex]
     \bottomrule
    \end{tabular}\vspace{-4mm}
\end{table}

\begin{table}[t]\vspace{5mm}
\centering
\caption{Simulation Parameters}\vspace{-1mm}
\begin{tabular}{c cc c} 
\toprule
Parameters & Value & Parameters & Value\\ [0.2ex] 
\midrule
$N$ & 100 & BS Antenna Spacing & $0.5 \lambda$ \\
$M$ & 64 & Central Frequency (GHz) & 28 \\
$K$ & 4 & LoS Path Loss (dB) & 56.9 + 22.0$\mathrm{lg}d$ \\
BS Location & (0 m, 0 m) & NLoS Path Loss (dB) & 60.3 + 36.7$\mathrm{lg}d$\\
\bottomrule
\end{tabular}\vspace{-3mm}
\label{tab:simulation parameters}
\end{table}

\begin{table}[t] \vspace{-0mm}
    \centering
    \caption{Computational Complexity}\vspace{0mm}
    \label{complexity}
    
    \begin{tabular}{c c c}
     \toprule
     No.&Method & Complexity \\
     \midrule
     1 & GMML& $\mathcal{O}(N_e N_o N_i K^2 M N)$ \\
     2 & GML & $\mathcal{O}(N_e N_o N_i (K^2 M N + M^2))$   \\
     3 & ML & $\mathcal{O}(N_e N_o N_i (K^2 M N + M^2))$ \\
     4 & DNN & $\mathcal{O}((N^2 + KMN + K^2M^2))$ \\
     5 & AO & $\mathcal{O}(L_3(L_1 K M^3 + L_2 K^2 N^2))$ \\ [1.0ex]
     \bottomrule
    \end{tabular}
    \vspace{-0mm}
\end{table}

\begin{figure}[t]\vspace{0mm}
	\begin{center}
		\centerline{\includegraphics[width=0.36\textwidth]{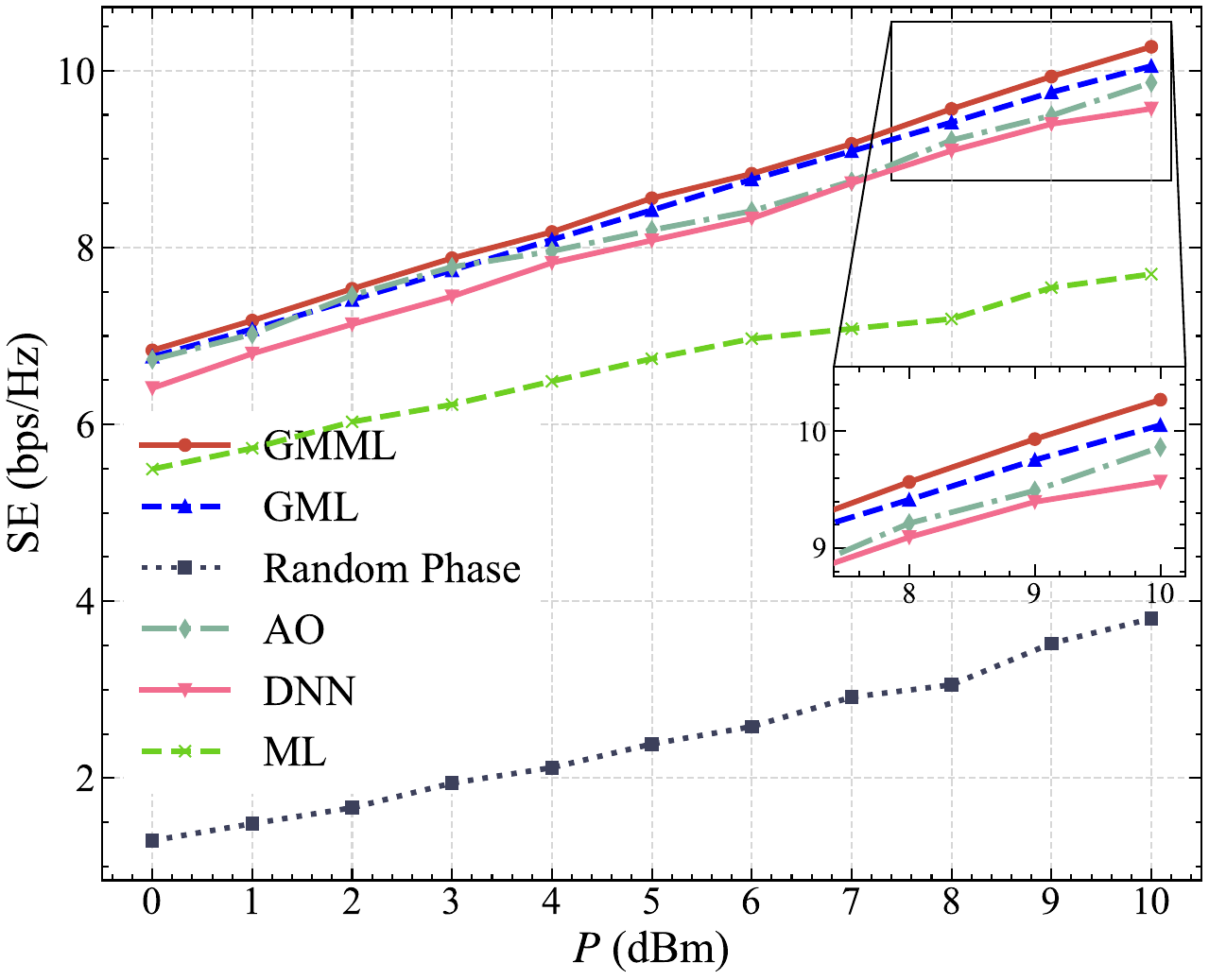}}  \vspace{-0mm}
		\captionsetup{font=footnotesize, name={Fig.}, labelsep=period} 
		\caption{\, SE vs. transmit power.}
		\label{fig:snr_perfect} \vspace{-8mm}
	\end{center}
\end{figure}

\begin{figure}[t]\vspace{0mm}
	\begin{center}
		\centerline{\includegraphics[width=0.36\textwidth]{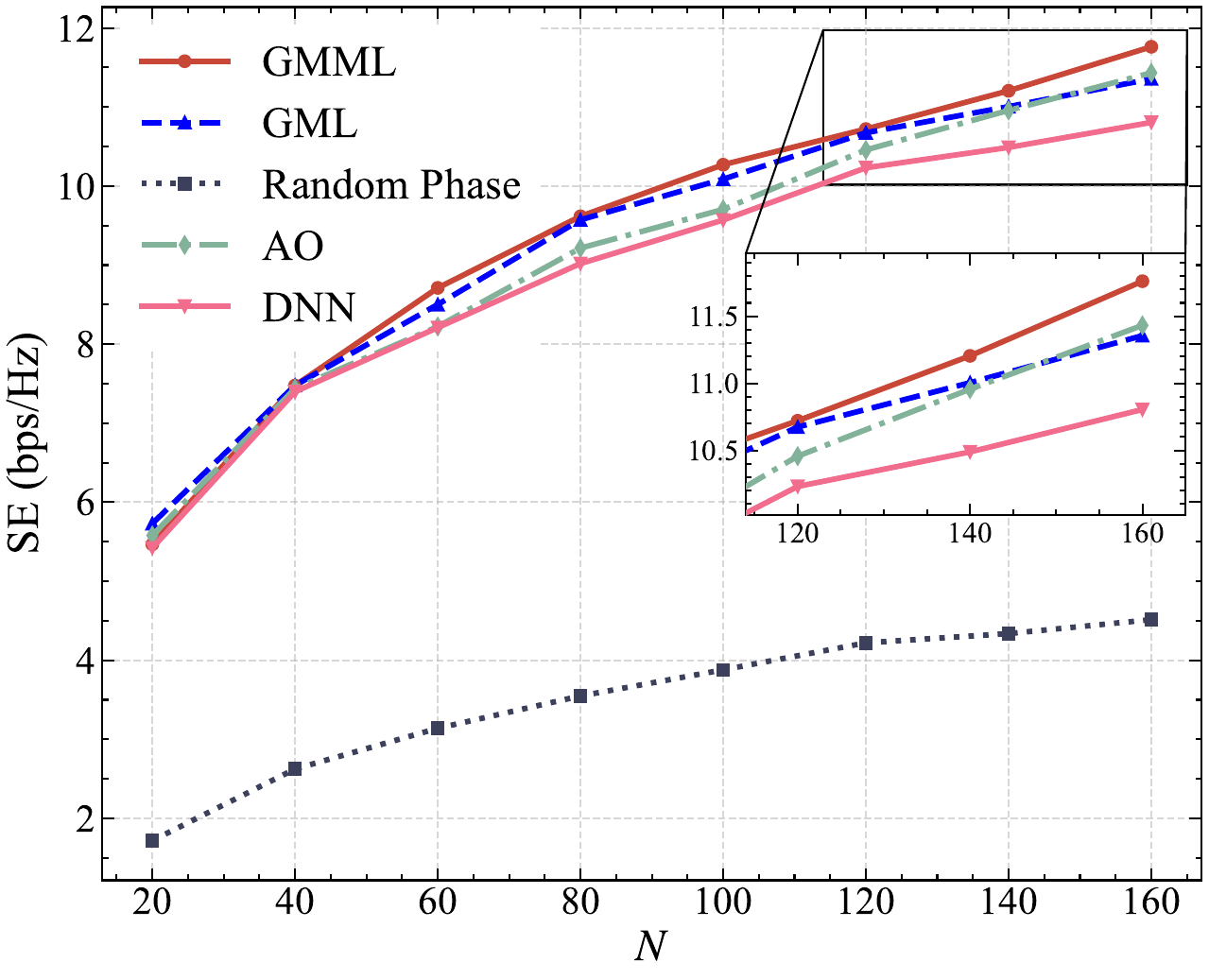}}  \vspace{-0mm}
		\captionsetup{font=footnotesize, name={Fig.}, labelsep=period} 
		\caption{\, SE vs. RIS element number.}
		\label{fig:ris_perfect} \vspace{-8mm}
	\end{center}
\end{figure}

\begin{figure}[t]\vspace{-0mm}
	\begin{center}
		\centerline{\includegraphics[width=0.36\textwidth]{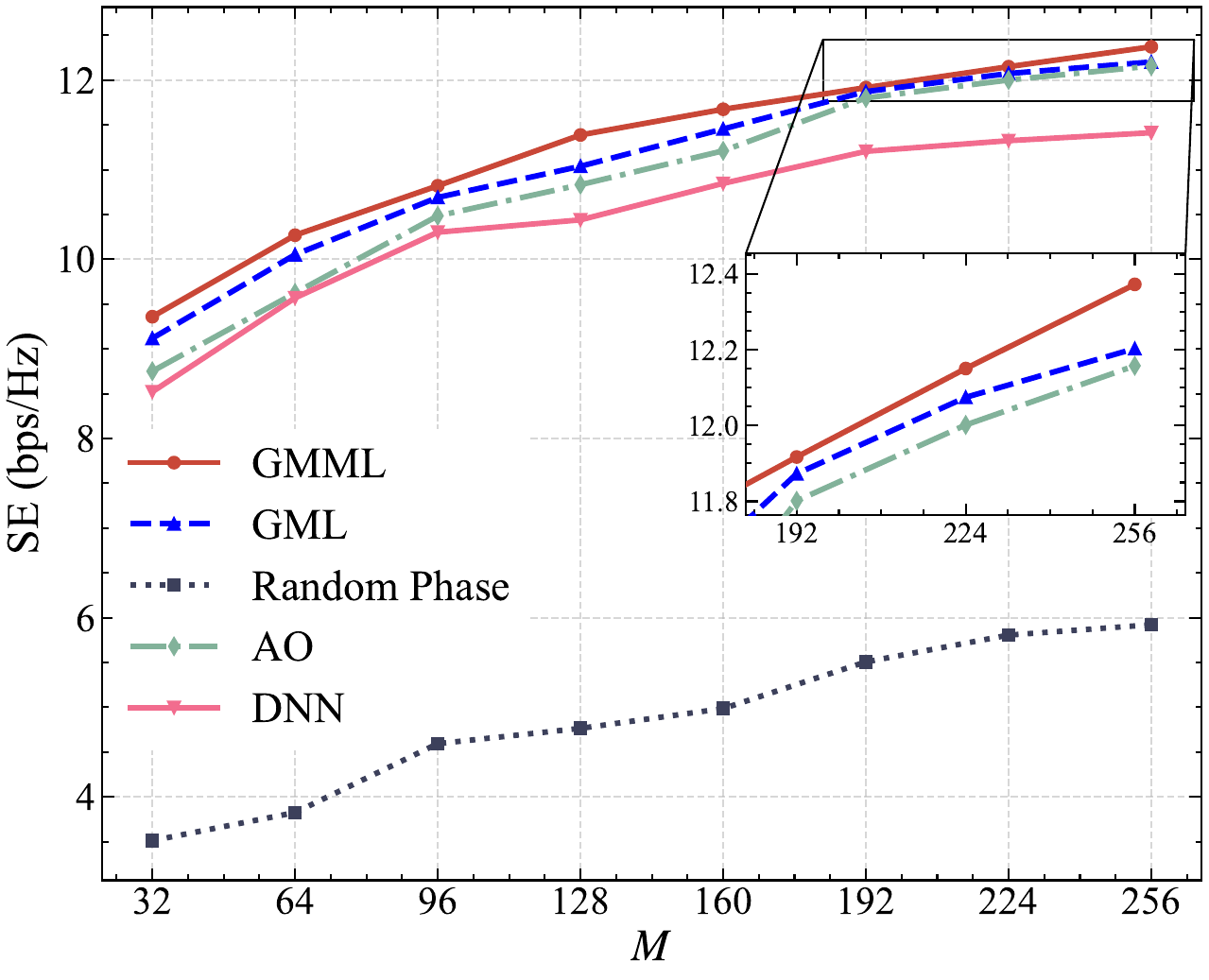}}  \vspace{-0mm}
		\captionsetup{font=footnotesize, name={Fig.}, labelsep=period} 
		\caption{\, SE vs. BS antenna number.}
		\label{fig:BS_perfect} \vspace{-8mm}
	\end{center}
\end{figure}

\begin{figure}[t]\vspace{0mm}
	\begin{center}
		\centerline{\includegraphics[width=0.36\textwidth]{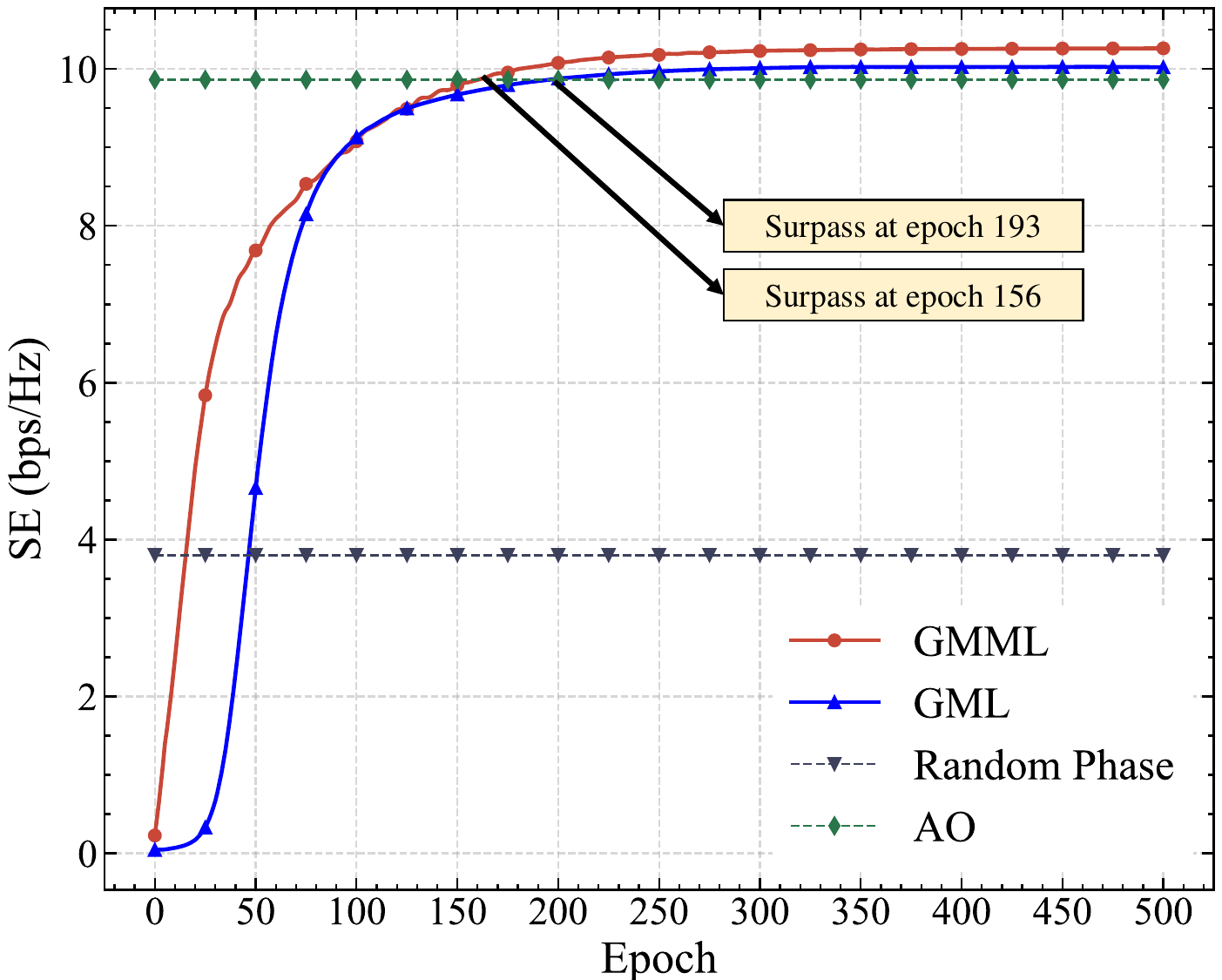}}  \vspace{-0mm}
		\captionsetup{font=footnotesize, name={Fig.}, labelsep=period} 
		\caption{\, SE vs. epoch number. }
		\label{fig:epoch_perfect} \vspace{-8mm}
	\end{center}
\end{figure}

\subsection{Perfect CSI Evaluation}
In this subsection, we evaluate the performance of the proposed scheme against baselines with perfect CSI. All the simulation curves have been averaged over $N_s$ independent channel realizations.
\par
In Fig. \ref{fig:snr_perfect}, we show the SE of the proposed algorithm and the baselines with respect to to $P$ when $M$ and $N$ are restricted to 64 and 100, respectively. All five methods exhibit an increase in SE when increasing $P$. Evidently, the approach with random phase significantly underperforms compared to the other four methods, as it only focuses on the optimization of the precoding matrix, while the other four schemes also optimize the RIS phase shifting matrix. This indicates the significance of optimizing the RIS phase shifting matrix in RIS-assisted communication scenarios. Besides, it is worth noting that the proposed GMML algorithm consistently outperforms other methods in the range of $0 \, \mathrm{dBm}$ to $10 \, \mathrm{dBm}$, and it achieves $4.11\%$ and $7.31\%$ higher performance than AO and DNN at $10 \, \mathrm{dBm}$. Moreover, GMML and GML exhibit noticeably superior performance over AO when $P$ exceeds $4 \, \mathrm{dBm}$. This can be attributed to the fact that the optimization space of the precoding matrix becomes larger and more complex as $P$ increases, resulting in an increasing number of local optima. As a result, the greedy AO method is more susceptible to becoming trapped, which explains why AO performs worse than the proposed meta learning based methods. Furthermore, among all the baseline methods except Random Phase, the ML method exhibits the poorest performance. This is because it is unable to capture the higher-order information in the input channel data without the gradient input mechanism. In addition, we approximate the upper bound of SE by running the AO scheme 100 times with independent random initialization and selecting the highest achieved SE for each sample as the upper bound, and GMML attains 90.5\% of the SE compared with the upper bound when $P$ is $10\mathrm{\,dBm}$.
\par
Next, in Fig. \ref{fig:ris_perfect}, $P$ is fixed to $10 \, \mathrm{dBm}$ and $M$ is fixed to 64, the achieved SE of the proposed algorithms and the baselines with respect to to $N$ is depicted. Thanks to the electromagnetic environment reconstruction ability of the RIS, the channel condition is enhanced as $N$ increases, resulting in improved performance of all schemes. When $N$ is small, GMML performs similarly to the baselines, as the cascaded channel condition is poor and there is limited room for joint optimization. However, as $N$ increases, GMML consistently outperforms other schemes. It can be observed that the AO method outperforms GML as $N$ exceeds 140. The reason for this phenomenon is that a larger search space of $\mathbf{\Theta}$ in GML would interference and slow down the optimization of $\mathbf{W}$, thereby reducing the overall rate. Moreover, the gap between GMML and DNN widens as $N$ increases, suggesting that GMML is better suited for RIS-aided communications with a large number of reflective elements than traditional deep learning methods.
\par
Fig. \ref{fig:BS_perfect} demonstrates the SE with $M$ when $N$ and $P$ are fixed to $100$ and $10 \, \mathrm{dBm}$. It is evident that all schemes benefit from a larger number of BS antennas. Notably, GMML achieves a $5.13\%$ higher performance than AO when $M=128$, but this advantage diminishes to $1.81\%$ as $M$ is increased to 256, which can be attributed to the undersized NNs. While the manifold learning helps to reduce the search space of the BS precoding matrix, it does not reduce the information within the optimization space. Therefore, enough neurons are required to memorize the characteristics of the optimization space. The performance of DNN remains the worst among the baseline methods, as the black-box DNN can not find the most efficient optimization path as the optimization space grows larger.
\begin{figure}\vspace{-0mm}
	\begin{center}
		\centerline{\includegraphics[width=0.36\textwidth]{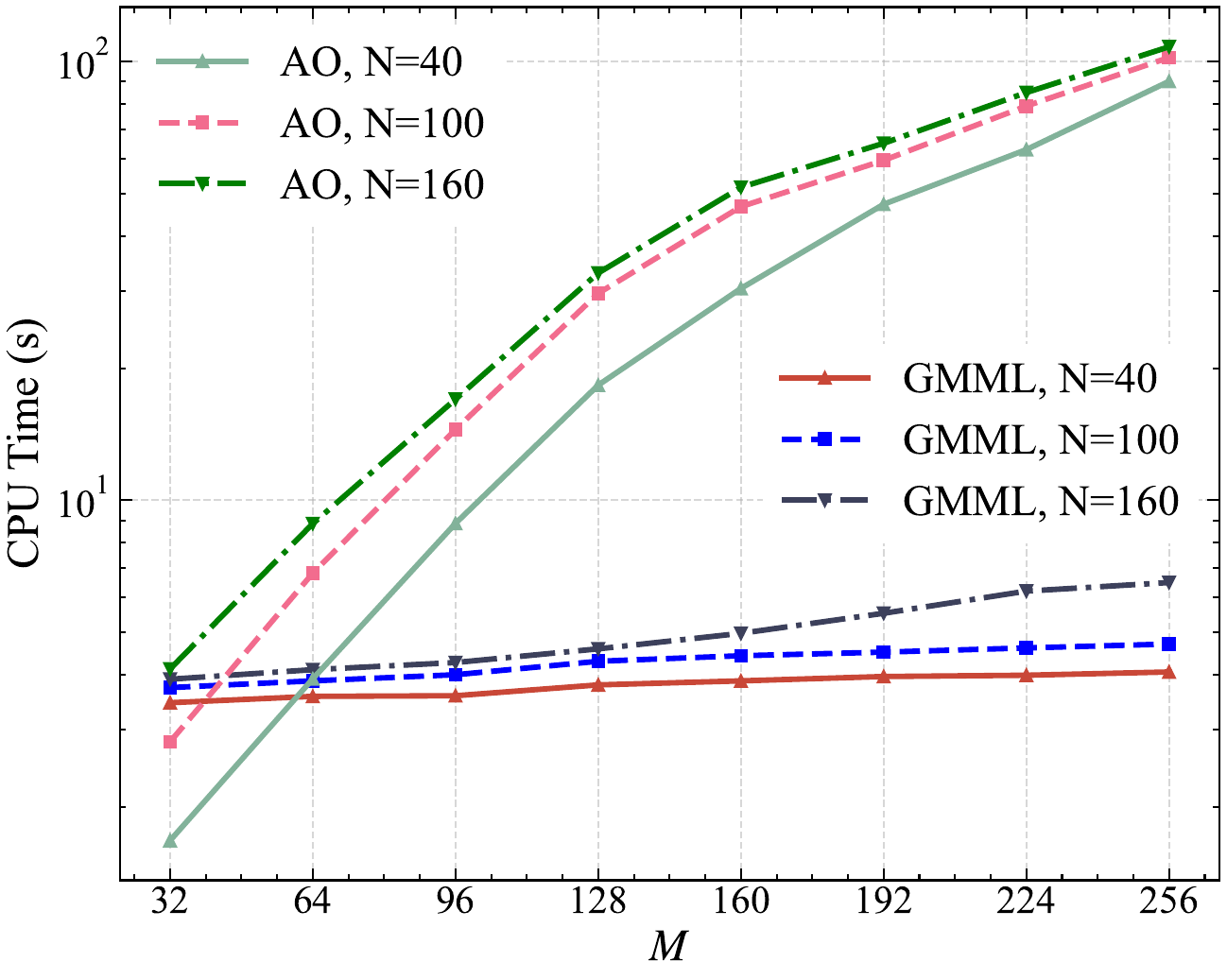}}  \vspace{-0mm}
		\captionsetup{font=footnotesize, name={Fig.}, labelsep=period} 
		\caption{\, Average CPU time vs. BS antenna and RIS element number.}
		\label{fig:time_RIS} \vspace{-8mm}
	\end{center}
\end{figure}

\begin{figure}[t]\vspace{0mm}
    \centering
    \begin{subfigure}[h]{0.36\textwidth}
        \centering
        \includegraphics[width=\textwidth]{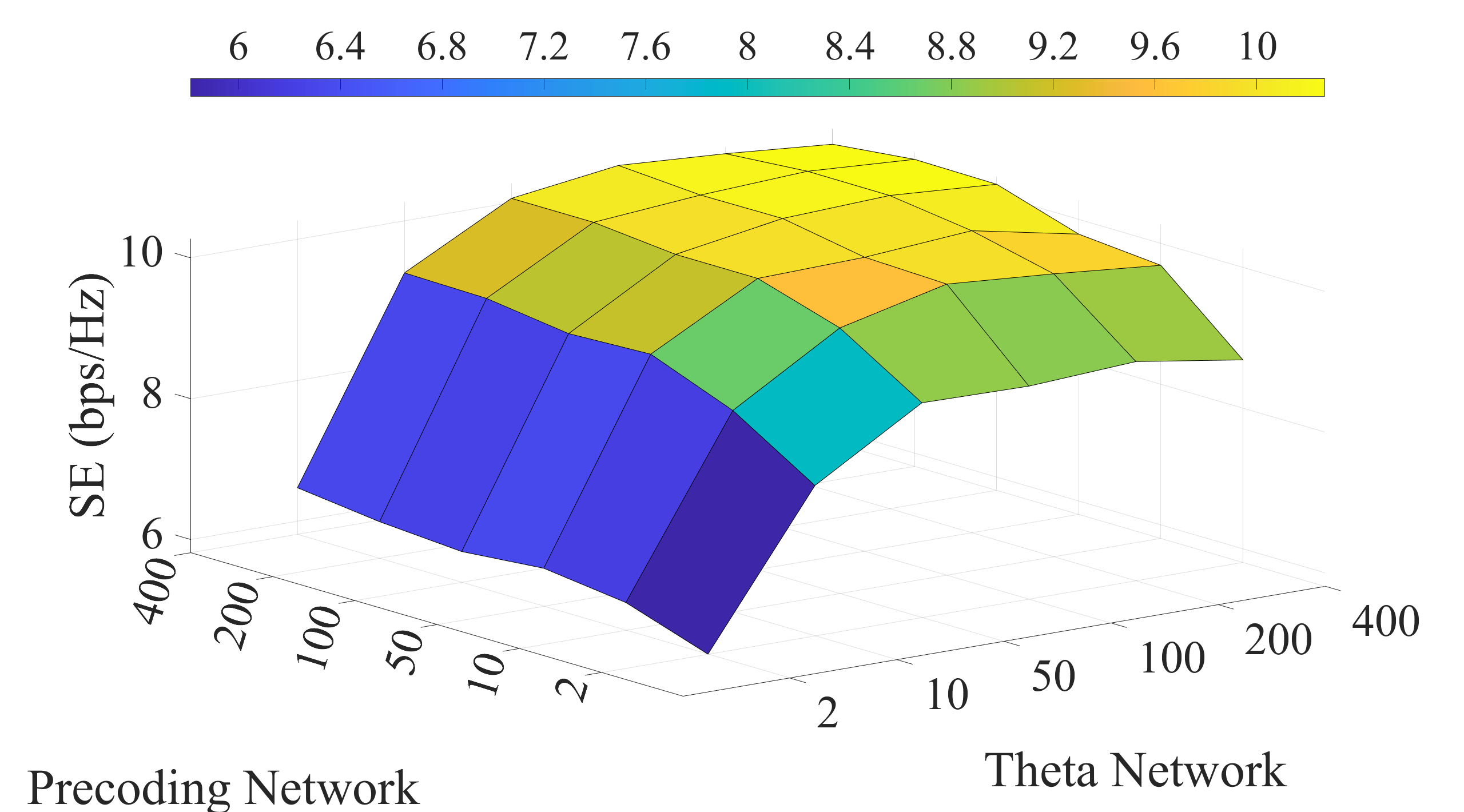}
        \captionsetup{font=footnotesize, name={Fig.}, labelsep=period} 
        \caption{}
        \label{fig:nn_perfect}
    \end{subfigure}
    \hfill
    \begin{subfigure}[h]{0.36\textwidth}
        \centering
        \includegraphics[width=\textwidth]{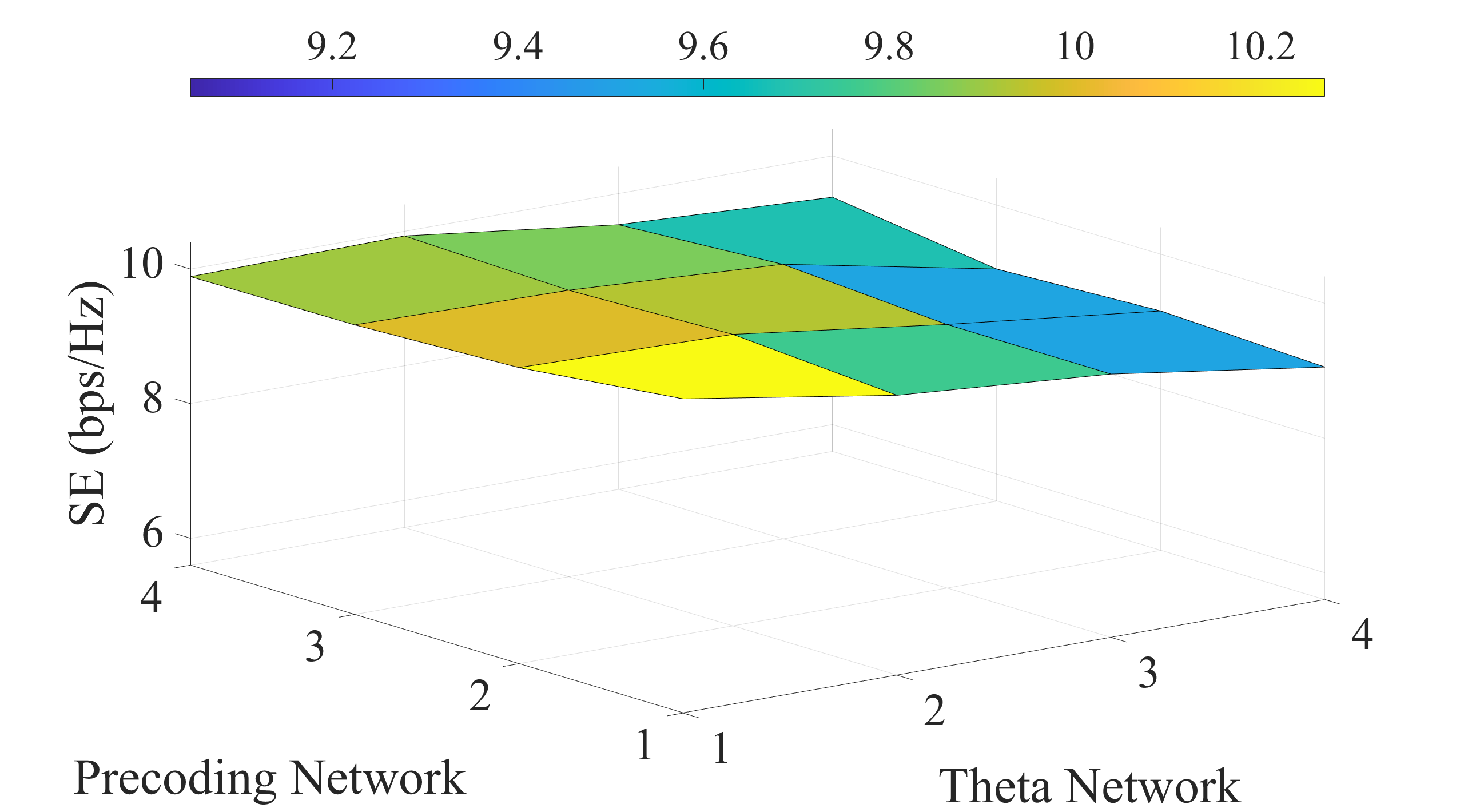}
        \captionsetup{font=footnotesize, name={Fig.}, labelsep=period} 
        \caption{}
        \label{fig:layer_perfect}
    \end{subfigure}
    \captionsetup{font=footnotesize, name={Fig.}, labelsep=period} 
    \caption{\, The SE performance comparison with different network structures: (a) SE vs. number of neurons per hidden layer and (b) SE vs. number of hidden layers.}
    \label{fig:nn_layer_perfect}\vspace{-3mm}
\end{figure}

\begin{figure}[t]\vspace{0mm}
	\begin{center}
		\centerline{\includegraphics[width=0.36\textwidth]{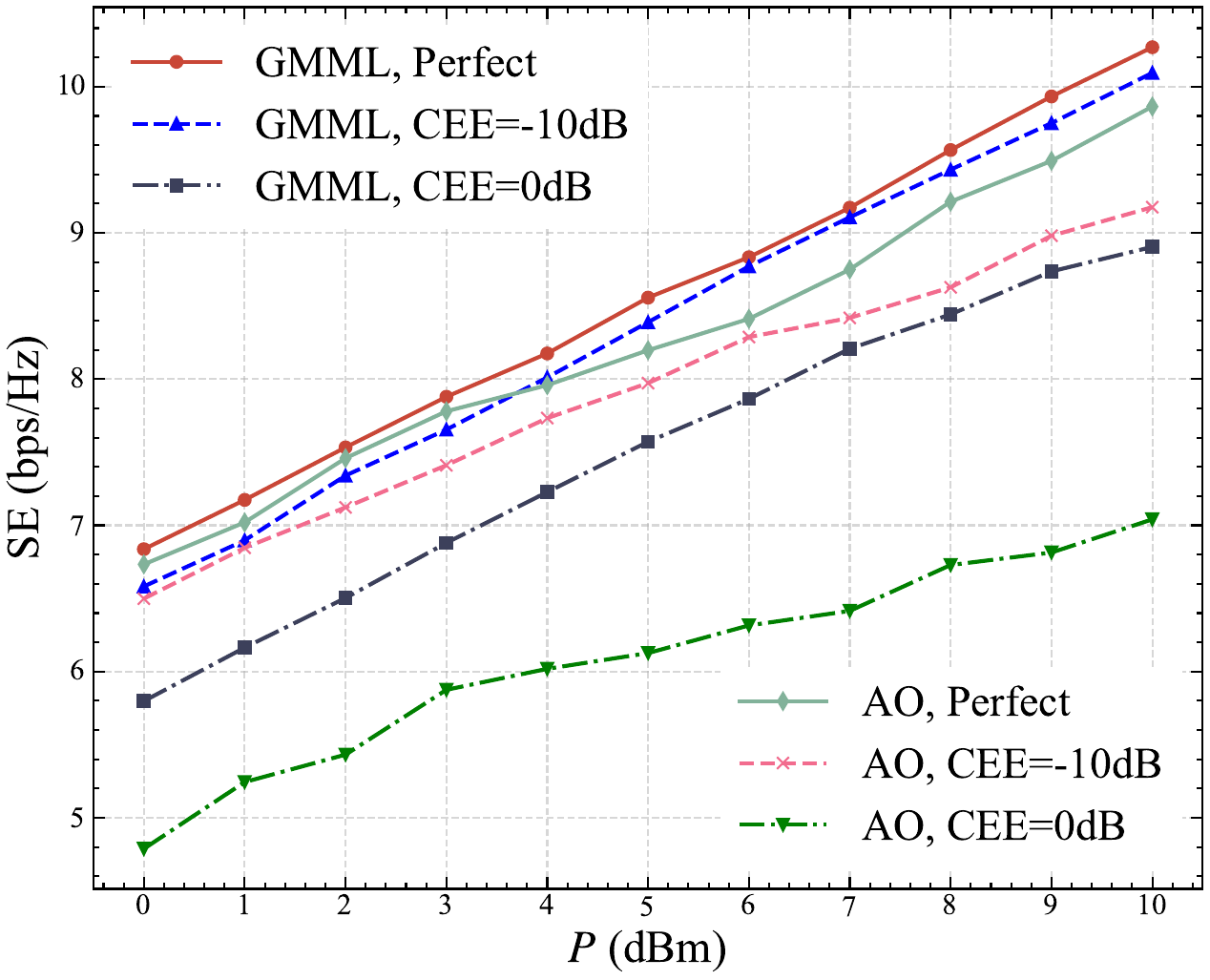}}  \vspace{-0mm}
		\captionsetup{font=footnotesize, name={Fig.}, labelsep=period} 
		\caption{\, SE performance with imperfect CSI.}
		\label{fig:cee_snr} \vspace{-8mm}
	\end{center}
\end{figure}

Then, Fig. \ref{fig:epoch_perfect} shows the convergence behavior of the proposed algorithms under the setting: $P=10 \, \mathrm{dBm}$, $M=64$ and $N=64$. It is noticeable that GMML and GML surpasses the convergence speed of AO at epoch 156 and epoch 193, respectively. Furthermore, it is obvious that the convergence speed of GMML is faster than GML. Specifically, GMML displays a steeper rise than GML during the initial epochs. This is due to the manifold learning approach in GMML which significantly reduces the target search space of the precoding matrix, thus making it easier to approach the optima and achieve higher performance.
\par
Afterwards, we compare the average CPU execution time of GMML and AO under different $M$ and $N$ while $P=10 \, \mathrm{dBm}$. It is noted that we adopt the logarithm of the CPU time here. As illustrated in Fig. \ref{fig:time_RIS}, both schemes exhibit an increase in average CPU time as $N$ increases while $M$ remains constant, and it is evident that the average CPU time of GMML increases at a slower rate than that of AO as $N$ increases from 40 to 100 and 160. Similarly, when $N$ is fixed, the average CPU time of GMML increases slowly with $M$ while AO exhibits a sharp rise. In particular, when $M=256$ and $N=160$, the AO algorithm takes $108.1$ seconds to converge, while the GMML algorithm only takes $4.7$ seconds. This suggests that our algorithm can speed up over $23$ times faster compared to the traditional AO, which matches the previous complexity analysis.
 
\par
Fig. \ref{fig:nn_layer_perfect}(a) and Fig. \ref{fig:nn_layer_perfect}(b) illustrate the effects of the number of neurons per hidden layer and the number of hidden layers on the SE. As shown in Fig. \ref{fig:nn_layer_perfect}(a), we can see that the SE rises with an increase in more neurons per layer for both networks, but reaches convergence when the number exceeds 100. This is because the NNs need an adequate number of neurons to store gradient information. If there are too few neurons per layer, the learning ability will deteriorate. However, it is noteworthy that Fig. \ref{fig:nn_layer_perfect}(b) indicates a different trend, where increasing the number of hidden layers leads to a degradation in performance. This could be attributed to inadequate training data for deeper networks, as the proposed algorithm is fed with the same single channel sample in each epoch.
\par

\begin{figure}[t]\vspace{-0mm}
	\begin{center}
		\centerline{\includegraphics[width=0.36\textwidth]{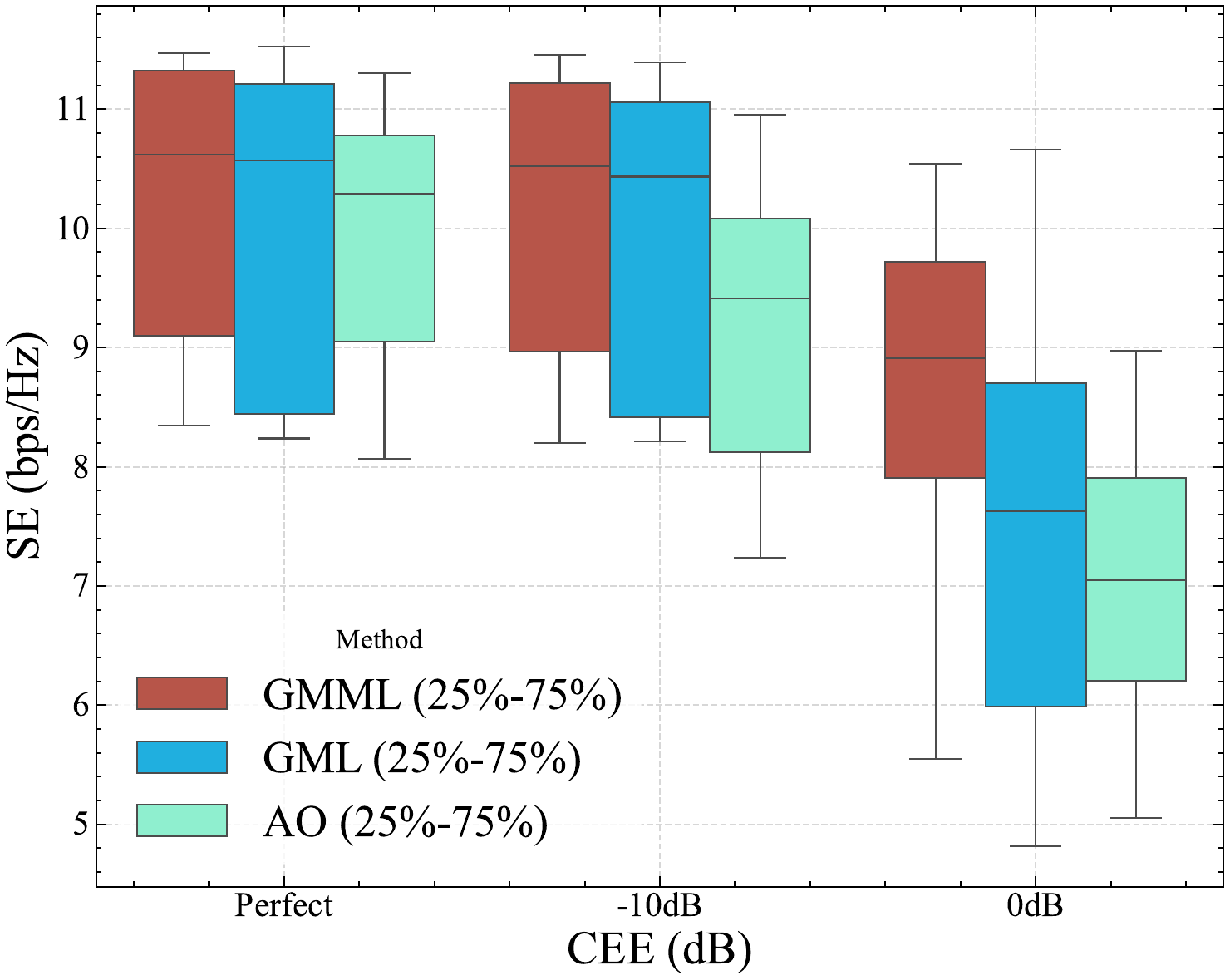}}  \vspace{-0mm}
		\captionsetup{font=footnotesize, name={Fig.}, labelsep=period} 
		\caption{\, Box plot of the variance of SE with imperfect CSI.}
		\label{fig:box_cee} \vspace{-10mm}
	\end{center}
\end{figure}

\begin{figure}[t]\vspace{6mm}
	\begin{center}
		\centerline{\includegraphics[width=0.36\textwidth]{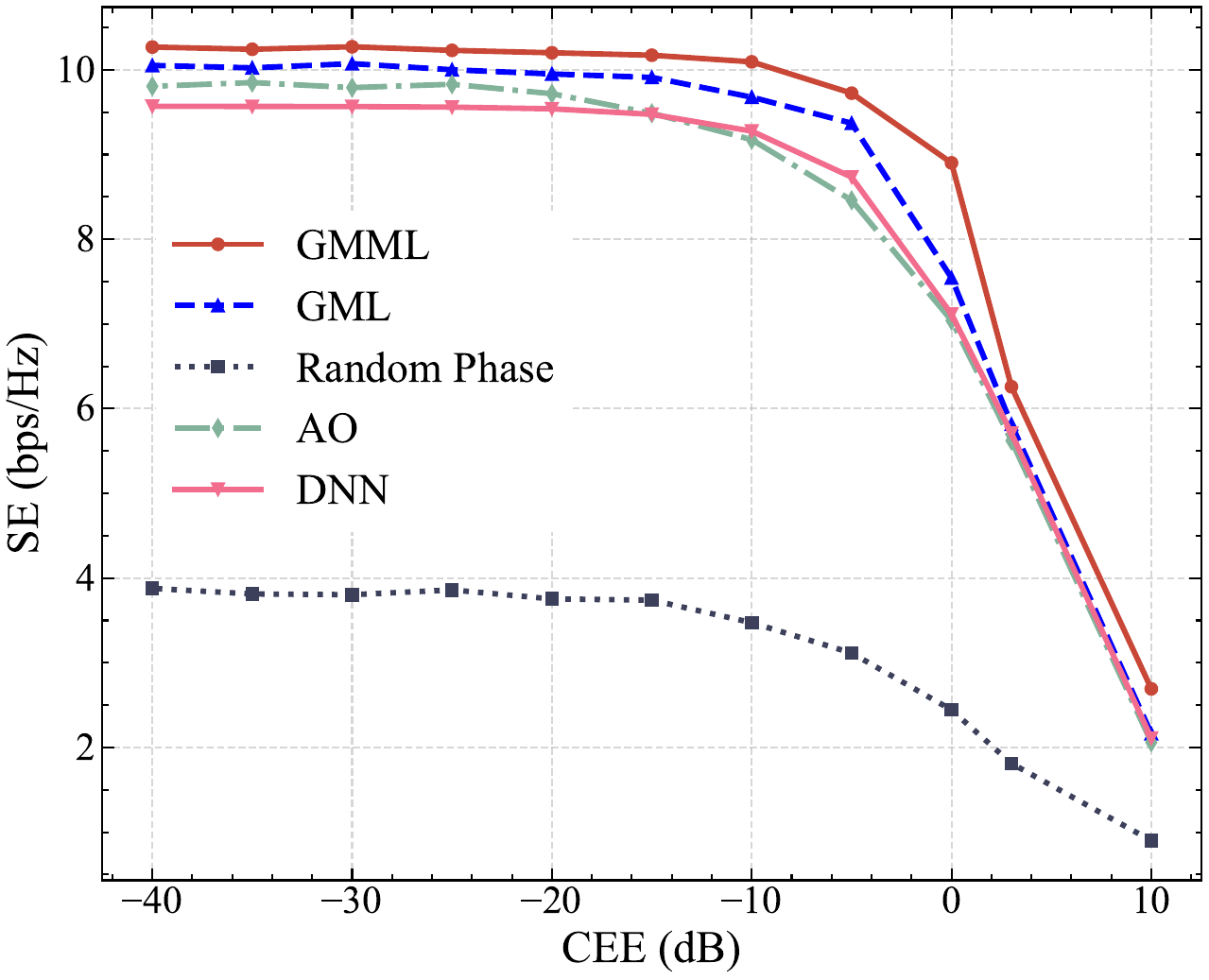}}  \vspace{-0mm}
		\captionsetup{font=footnotesize, name={Fig.}, labelsep=period} 
		\caption{\, SE vs. CEE when $P=10 \, \mathrm{dBm}$.}
		\label{fig:SE_cee} \vspace{-8mm}
	\end{center}
\end{figure}

\begin{figure}[t]\vspace{-0mm}
	\begin{center}
		\centerline{\includegraphics[width=0.36\textwidth]{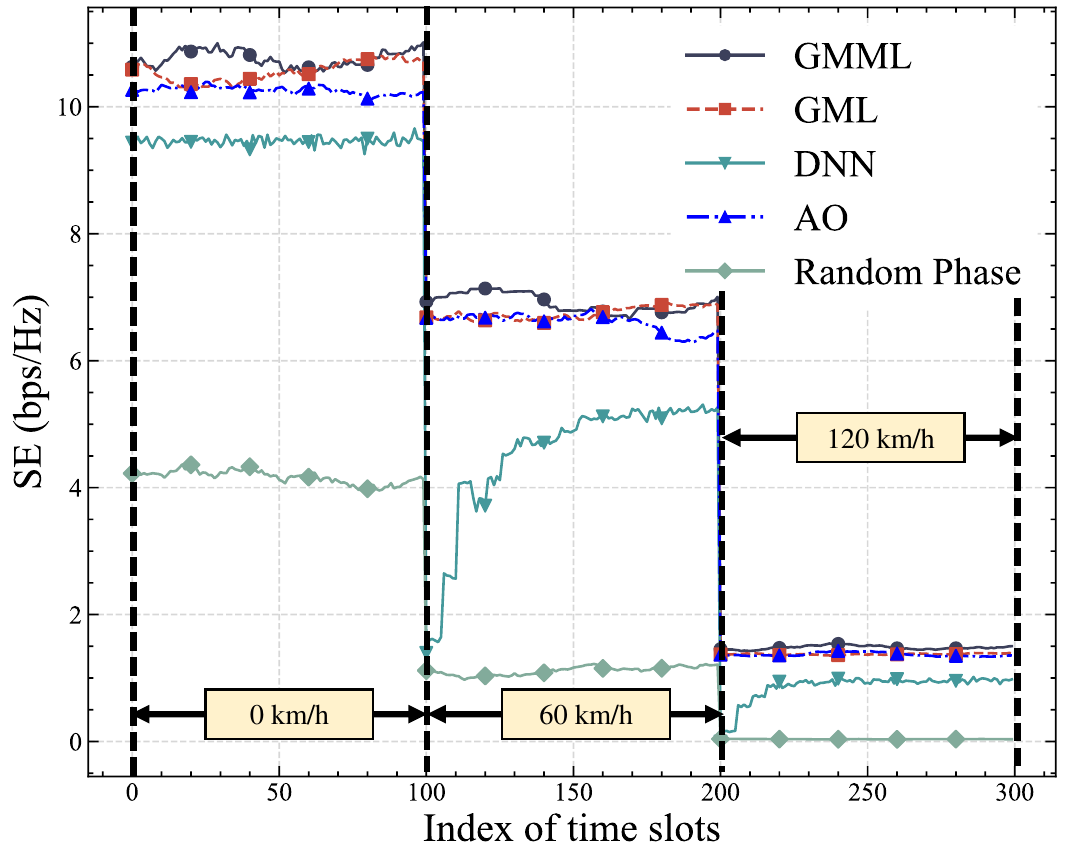}}  \vspace{-0mm}
		\captionsetup{font=footnotesize, name={Fig.}, labelsep=period}
		\caption{\, Performance comparison in dynamic scenarios.}
		\label{fig:dynamic} \vspace{-8mm}
	\end{center}
\end{figure}

\subsection{Imperfect CSI Evaluation}
In this subsection, we evaluate the performance of the proposed scheme against baselines with imperfect CSI. The CEEs in the simulations refer to the estimation bias of both $\mathbf{G}$ and $\mathbf{H}$. All the simulation curves have been averaged over $N_s$ independent channel realizations.
\par
Fig. \ref{fig:cee_snr} compares the SE in imperfect CSI with different $P$ and CEE, where the system setting is the same as Fig. \ref{fig:snr_perfect}. GMML demonstrates stronger robustness against imperfect CSI than the AO method, particularly when channel estimation accuracy is low and $P$ is high. To be specific, when $P=10 \, \mathrm{dBm}$, compared to the perfect CSI setup, the proposed GMML algorithm can achieve the performance of $98.29 \%$ and $86.70 \%$ with the CEE at $-10 \, \mathrm{dB}$ and $0 \, \mathrm{dB}$, respectively. In contrast, the traditional AO algorithm attains only the performance of $93.02 \%$ and $71.38 \%$ in an imperfect CSI setup with corresponding CEE values. In Fig. \ref{fig:box_cee}, we demonstrate the SE performance and variance of the GMML and the baselines when $P=10 \, \mathrm{dBm}$. It is apparent that meta learning based methods exhibit higher average performance than AO algorithm, while GMML demonstrates smaller variance in contrast to GML, but similar variance to AO. This indicates GMML can achieve high performance while maintaining stability and consistency in practical imperfect CSI setup.
\par
Next, we observe the corresponding SE while changing CEE in Fig. \ref{fig:SE_cee} with $P$ set to 10 dBm. It is evident that a larger CEE leads to a decrease in the SE for all schemes. Among all imperfect channel setups with different CEE, GMML exhibits the best performance. This implies that GMML can handle greater channel uncertainty while achieving the same SE. As a result, the need for accurate channel estimation is greatly reduced, consequently decreasing the overall system overhead. The proposed method is robust to imperfect CSI for two main reasons. First, the optimization targets are not individual variables but the entire search trajectory. This meta-learning technique enables the network to learn optimization strategies, allowing it to dynamically select better trajectories to filter noise and minimize the impact of channel estimation errors on overall performance. Second, we apply the gradient input mechanism. This approach may smooth the original highly non-convex optimization problem, reducing the impact of channel estimation errors in the gradient space. As a result, the optimization trajectory is less likely to be degraded by these errors.

\subsection{Dynamic Scenario Evaluation}
In this subsection, we evaluate the adaptability of the proposed scheme against baselines in dynamic scenarios. It is assumed that the mobile users transition from outdoor to urban and then to freeway, moving at speeds of 0 km/h, 60 km/h, and 120 km/h, respectively. Technical standards \cite{3gpp.36.814} and \cite{3gpp.36.885} are separately used to generate perfect channel data for stationary and moving scenarios.
\par
Fig. \ref{fig:dynamic} demonstrates the adaptation performance of the proposed scheme against the baselines across three different scenarios. Each scenario lasts for 100 time slots, with five adaptation channels and five testing channels in each time slot, and the adaptation channels are exclusively for DNN. It is observed that GMML still outperforms other schemes in terms of average SE, while the DNN based scheme has much worse performance. When the users move from outdoor to urban, the gap between DNN and other schemes widens. This is because the DNN still applies knowledge from the previous scenario, which does not match the characteristics of the new environment. Furthermore, compared to other schemes, DNN experiences a significant decrease in performance when the communication scenario changes. This can be attributed to the limited generalization capacity and strong memory effect of DNN when applied to numerical optimization problems. Adequate adaptation data and a considerable amount of time are required for DNN to adjust to new environments. It is noted that the proposed GMML and its simplified version GML work similarly to the offline algorithm AO, in which the embedded NNs are updated within each channel optimization, but are randomly initialized across each channel realization. Therefore, GMML handles new scenarios without requiring any additional fine-tuning.


\section{Conclusion}\label{sec:conclusion}
In this paper, a pre-training free scheme for RIS-aided communications named GMML was proposed. To jointly optimize the precoding matrix and the phase shifting matrix at the BS and the RIS, respectively, GMML fuses the meta learning and manifold learning, which could improve the overall SE and reduce the overhead of the high-dimensional signal process.
Extensive numerical simulations demonstrate that not only can the proposed scheme outperform baseline methods in scenarios with perfect CSI, but also it can keep the strong robustness in scenarios with imperfect SCI. Moreover, results also show the proposed scheme exhibits the superior adaptability in dynamic scenarios. In the future, we will focus on exploring the beamforming in RISs-aided communications with incomplete CSI or even no CSI.


\begin{appendices}
\vspace{2mm}
\section{Proof of Proposition 1}\label{appendix1}
For any nontrivial stationary point $\mathbf{w}_k^*$, we can employ the Karush-Kuhn-Tucker (KKT) condition on the original system SE maximization problem \eqref{optimization problem} and derive:
\begin{subequations}\label{KKT}
\begin{align}
& \sum_{i=1}^{K}\omega_i \nabla_{\mathbf{w}_k}R_i-\lambda^*\mathbf{w}_k^*=0, \label{KKT1} \\
& \left(\sum_{k=1}^{K}\mathrm{Tr}(\mathbf{w}_k^*(\mathbf{w}_k^*)^H)-P\right) \cdot \lambda^* = 0, \label{KKT2} \\
& \sum_{k=1}^{K}\mathrm{Tr}(\mathbf{w}_k^*(\mathbf{w}_k^*)^H) \leq P, \label{KKT3} \\
& \lambda^* \geq 0, \label{KKT4}
\end{align}
\end{subequations}
where $\lambda^*$ is the Lagrange multiplier.
\par
Now we would prove by contradiction that $\lambda^*$ is strictly positive. First we take the derivative of $R_k$ with respective to $\mathbf{w}_k$ and yields
\begin{align}\label{devrk}
    & \nabla_{\mathbf{w}_k}R_k \nonumber \\
    &= \nabla_{\mathbf{w}_k} \log_2(1 + \frac{|\mathbf{h}_k^H\mathbf{\Theta G w}_k|^2}{\sigma^2 + \sum_{j \neq k}^K|\mathbf{h}_k^H\mathbf{\Theta G w}_j|^2}) \nonumber \\
    & = \nabla_{\mathbf{w}_k}  \log_2(\frac{\sigma^2 + \sum_{j = 1}^K|\mathbf{h}_k^H\mathbf{\Theta G w}_j|^2}{\sigma^2 + \sum_{j \neq k}^K|\mathbf{h}_k^H\mathbf{\Theta G w}_j|^2}) \nonumber \\
    & = \nabla_{\mathbf{w}_k} \log_2(\sigma^2 + \sum_{j = 1}^K|\mathbf{h}_k^H\mathbf{\Theta G w}_j|^2) \nonumber \\
    & - \nabla_{\mathbf{w}_k} \log_2(\sigma^2 + \sum_{j \neq k}^K|\mathbf{h}_k^H\mathbf{\Theta G w}_j|^2)  \nonumber \\
    & = \nabla_{\mathbf{w}_k}  \log_2(\sigma^2 + \sum_{j = 1}^K|\mathbf{h}_k^H\mathbf{\Theta G w}_j|^2)  \nonumber \\
    & = \frac{1}{\ln{2}}(\sigma^2 + \sum_{j = 1}^K|\mathbf{h}_k^H\mathbf{\Theta G w}_j|^2)^{-1} \cdot \nabla_{\mathbf{w}_k} |\mathbf{h}_k^H\mathbf{\Theta G w}_k|^2 \nonumber \\
    & = \frac{2}{\ln{2}}(\mathbf{h}_k^H\mathbf{\Theta G })^H \cdot \frac{(\mathbf{h}_k^H\mathbf{\Theta G w}_k)}{(\sigma^2 + \sum_{j = 1}^K|\mathbf{h}_k^H\mathbf{\Theta G w}_j|^2)^{-1}} \nonumber \\
    & = \frac{2}{\ln{2}}(\mathbf{h}_k^H\mathbf{\Theta G })^H \cdot {z}_{kk}, \nonumber \\
\end{align}
in which 
\begin{equation}
    {z}_{kk}= \frac{(\mathbf{h}_k^H\mathbf{\Theta G w}_k)}{(\sigma^2 + \sum_{j = 1}^K|\mathbf{h}_k^H\mathbf{\Theta G w}_j|^2)^{-1}}.
\end{equation}
Similarly, we can get
\begin{align}\label{devri}
    & \nabla_{\mathbf{w}_k}R_i \nonumber \\
    & = \frac{2}{\ln{2}}(\mathbf{h}_i^H\mathbf{\Theta G })^H \cdot \frac{(\mathbf{h}_i^H\mathbf{\Theta G w}_k)}{(\sigma^2 + \sum_{j = 1}^K|\mathbf{h}_i^H\mathbf{\Theta G w}_j|^2)^{-1}} \nonumber \\
    & - \frac{2}{\ln{2}}(\mathbf{h}_i^H\mathbf{\Theta G })^H \cdot \frac{(\mathbf{h}_i^H\mathbf{\Theta G w}_k)}{(\sigma^2 + \sum_{j \neq i}^K|\mathbf{h}_i^H\mathbf{\Theta G w}_j|^2)^{-1}} \nonumber \\
    & = \frac{2}{\ln{2}}(\mathbf{h}_i^H\mathbf{\Theta G })^H \cdot {z}_{ik}, \nonumber \\
\end{align}
in which
\begin{align}
    {z}_{ik} & = \frac{(\mathbf{h}_i^H\mathbf{\Theta G w}_k)}{(\sigma^2 + \sum_{j = 1}^K|\mathbf{h}_i^H\mathbf{\Theta G w}_j|^2)^{-1}} \nonumber
      \\
    & - \frac{(\mathbf{h}_i^H\mathbf{\Theta G w}_k)}{(\sigma^2 + \sum_{j \neq i}^K|\mathbf{h}_i^H\mathbf{\Theta G w}_j|^2)^{-1}}.
\end{align}
If $\lambda^*$ equals 0, we can substitute \eqref{devrk} and \eqref{devri} into \eqref{KKT1}, and multiply it with $\mathbf{w}_k^*$, which derives
\begin{equation}\label{contradict}
\begin{split}
    & \sum_{k=1}^{K}\omega_k\left(\sum_{j=1}^{K}\mathbf{h}_k^H\mathbf{\Theta G }\mathbf{w}_j^*(\mathbf{w}_j^*)^H(\mathbf{h}_k^H\mathbf{\Theta G })^H + \sigma^2\right)^{-1} \\
    & = \sum_{k=1}^{K}\omega_k\left(\sum_{j\neq k}^{K}\mathbf{h}_k^H\mathbf{\Theta G }\mathbf{w}_j^*(\mathbf{w}_j^*)^H(\mathbf{h}_k^H\mathbf{\Theta G })^H + \sigma^2\right)^{-1}, \\\end{split}   
\end{equation}
which indicates a trivial result: $\mathbf{h}_k^H\mathbf{\Theta G}\mathbf{w}_j^*=0$. This contradicts the initial assumption that $\mathbf{w}_j^*$ is a nontrivial stationary point. Given that $\lambda^* \geq 0$, $\lambda^*$ will indeed be strictly positive for any nontrivial stationary point $\mathbf{w}_j^*$. Consequently, this leads to the conclusion that $\sum_{k=1}^{K}\mathrm{Tr}(\mathbf{w}_k^*(\mathbf{w}_k^*)^H) = P$ as derived from \eqref{KKT2}. This completes the proof of \textit{Proposition 1}. It is noted that although $\mathbf{\Theta}$ would change frequently in the whole meta learning optimization procedure, this proposition still holds.
\vspace{2mm}
\section{Proof of Proposition 2}\label{appendix2}
Substituting \eqref{devrk} and \eqref{devri} into \eqref{KKT1} yields
\begin{equation}\label{rangespace}
\begin{split}
     \mathbf{w}_k^*
    &= \frac{2}{\lambda^*} \left(\omega_k (\mathbf{h}_k^H\mathbf{\Theta G})^H {z}_{kk} + \sum_{i \neq k}^{K}\omega_i (\mathbf{h}_k^H\mathbf{\Theta G})^H {z}_{ik}\right) \\
    & = \mathbf{H}_c^H \mathbf{X}_k,
\end{split}   
\end{equation}
in which $\mathbf{H}_c=[\mathbf{h}_{c,1}^H, \mathbf{h}_{c,2}^H,\cdots,\mathbf{h}_{c, K}^H]^H \in \mathbb{C}^{K \times M}$
and $\mathbf{X}_k=\frac{2}{\lambda^*}[\omega_1 z_{1k}, \omega_2 z_{2k}, \cdots,\omega_k z_{kk},\cdots, \omega_K z_{Kk}]^T\in \mathbb{C}^{K \times 1}$. Thus we complete the proof of \textit{Proposition 2}. 
\vspace{-2mm}
\section{GMML Complexity Analysis}\label{appendix3}
The overall complexity of GMML can be analyzed as follows. First, we analyze the complexity within an inner iteration.
\par
The analysis starts by examining the complexity of the \textit{Theta Network}. The primary contributors to this complexity are the computations involved in calculating SE and the NNs. To calculate SE, we primarily rely on equations \eqref{snr single user} and \eqref{spectrum efficiency}. Considering that the dimension of $\mathbf{h}_k^H$ is $N \times 1$ and $\mathbf{\Theta}$ is a diagonal matrix, the computational cost for multiplying $\mathbf{h}_k^H$ with $\mathbf{\Theta}$ is $\mathcal{O}(N)$.
The matrix $\mathbf{G}$ has a dimension of $N \times M$, adding a complexity of $\mathcal{O}(MN)$ when multiplied, while the multiplication involving $\mathbf{w}_j$ contributes an additional $\mathcal{O}(M)$. Therefore, the total complexity for computing $\mathbf{h}_k^H\mathbf{\Theta G w}_j$ is $\mathcal{O}(N + MN + M) = \mathcal{O}(MN)$. This calculation is performed $K$ times for computing the SINR, and subsequently the computation of SE involves calculating SINR $K$ times. Thus, the overall complexity for computing SE amounts to $\mathcal{O}(K^2 MN)$. Owing to the automatic differentiation, the computation of the gradient for $\mathbf{\Theta}$ is carried out concurrently, not contributing to the total complexity. Since the embedded DNNs are relatively small and shallow, the computational complexity of the NNs is approximately $\mathcal{O}(N)$. Therefore, the complexity in a single iteration within the \textit{Theta Network} can be expressed as $\mathcal{O}(K^2 NM) + \mathcal{O}(N) = \mathcal{O}(K^2 NM)$.
\par
The computational complexity of the \textit{Precoding Network} is similar to that of the \textit{Theta Network}, with an additional complexity owing to the computation of power normalization \eqref{antenna_power_constraint} and precoding matrix recovery \eqref{precoding_matrix_recovery}, whose complexity are both $\mathcal{O}(MK^2)$. The complexity involved in calculating the SE remains $\mathcal{O}(K^2 NM)$. The complexity related to the DNN operations is $\mathcal{O}(K^2)$. When these individual complexities are combined, the overall computational complexity of the \textit{Precoding Network} in a single iteration is $\mathcal{O}(MK^2) + \mathcal{O}(K^2) + \mathcal{O}(K^2 NM)=\mathcal{O}(K^2 MN)$.
\par
Considering the number of the inner, outer and epoch iterations, the overall complexity of the proposed GMML algorithm is 
$N_e N_o N_i(\mathcal{O}(K^2 M N) + \mathcal{O}(K^2 M N)) = \mathcal{O}(N_e N_o N_i K^2 M N)$.
\end{appendices}


\small
\bibliographystyle{IEEEtran}
\bibliography{bib}

\begin{thebibliography}{10}
\providecommand{\url}[1]{#1}
\csname url@samestyle\endcsname
\providecommand{\newblock}{\relax}
\providecommand{\bibinfo}[2]{#2}
\providecommand{\BIBentrySTDinterwordspacing}{\spaceskip=0pt\relax}
\providecommand{\BIBentryALTinterwordstretchfactor}{4}
\providecommand{\BIBentryALTinterwordspacing}{\spaceskip=\fontdimen2\font plus
\BIBentryALTinterwordstretchfactor\fontdimen3\font minus
  \fontdimen4\font\relax}
\providecommand{\BIBforeignlanguage}[2]{{%
\expandafter\ifx\csname l@#1\endcsname\relax
\typeout{** WARNING: IEEEtran.bst: No hyphenation pattern has been}%
\typeout{** loaded for the language `#1'. Using the pattern for}%
\typeout{** the default language instead.}%
\else
\language=\csname l@#1\endcsname
\fi
#2}}
\providecommand{\BIBdecl}{\relax}
\BIBdecl

\bibitem{hcwRIS}
C.~Huang, A.~Zappone, G.~C. Alexandropoulos, M.~Debbah, and C.~Yuen,
  ``{Reconfigurable Intelligent Surfaces for Energy Efficiency in Wireless
  Communication},'' \emph{IEEE Trans. Wireless Commun.}, vol.~18, no.~8, pp.
  4157--4170, Aug. 2019.

\bibitem{hcwCE}
L.~Wei, C.~Huang, G.~C. Alexandropoulos, C.~Yuen, Z.~Zhang, and M.~Debbah,
  ``{Channel Estimation for RIS-Empowered Multi-User MISO Wireless
  Communications},'' \emph{IEEE Trans. Commun.}, vol.~69, no.~6, pp.
  4144--4157, Jun. 2021.

\bibitem{wqqRIS}
Q.~Wu and R.~Zhang, ``{Intelligent Reflecting Surface Enhanced Wireless Network
  via Joint Active and Passive Beamforming},'' \emph{IEEE Trans. Wireless
  Commun.}, vol.~18, no.~11, pp. 5394--5409, Nov. 2019.

\bibitem{hcwRL}
C.~Huang, Z.~Yang, G.~C. Alexandropoulos, K.~Xiong, L.~Wei, C.~Yuen, Z.~Zhang,
  and M.~Debbah, ``{Multi-Hop RIS-Empowered Terahertz Communications: A
  DRL-Based Hybrid Beamforming Design},'' \emph{IEEE J. Sel. Areas Commun.},
  vol.~39, no.~6, pp. 1663--1677, Jun. 2021.

\bibitem{lywRIS}
Y.~Liu, X.~Liu, X.~Mu, T.~Hou, J.~Xu, M.~Di~Renzo, and N.~Al-Dhahir,
  ``{Reconfigurable Intelligent Surfaces: Principles and Opportunities},''
  \emph{IEEE Commun. Surv. Tutor.}, vol.~23, no.~3, pp. 1546--1577, 2021.

\bibitem{ganxuRIS1}
X.~Gan, C.~Zhong, C.~Huang, Z.~Yang, and Z.~Zhang, ``{Multiple RISs Assisted
  Cell-Free Networks With Two-Timescale CSI: Performance Analysis and System
  Design},'' \emph{IEEE Trans. Commun.}, vol.~70, no.~11, pp. 7696--7710, Nov.
  2022.

\bibitem{wqqRIS3}
S.~Abeywickrama, R.~Zhang, Q.~Wu, and C.~Yuen, ``{Intelligent Reflecting
  Surface: Practical Phase Shift Model and Beamforming Optimization},''
  \emph{IEEE Trans. Commun.}, vol.~68, no.~9, pp. 5849--5863, Sept. 2020.

\bibitem{robust1}
Z.~Chen, J.~Tang, X.~Y. Zhang, Q.~Wu, G.~Chen, and K.-K. Wong, ``{Robust Hybrid
  Beamforming Design for Multi-RIS Assisted MIMO System With Imperfect CSI},''
  \emph{IEEE Trans. Wireless Commun.}, vol.~22, no.~6, pp. 3913--3926, Jun.
  2023.

\bibitem{robust2}
M.~Gao, J.~Yang, H.~Li, and Y.~Wang, ``{Robust Beamforming Optimization Design
  for RIS-Aided MIMO Systems With Practical Phase Shift Model and Imperfect
  CSI},'' \emph{IEEE Internet Things J.}, vol.~11, no.~1, pp. 958--973, Jan.
  2024.

\bibitem{guohuayan}
H.~Guo, Y.-C. Liang, J.~Chen, and E.~G. Larsson, ``{Weighted Sum-Rate
  Maximization for Reconfigurable Intelligent Surface Aided Wireless
  Networks},'' \emph{IEEE Trans. Wireless Commun.}, vol.~19, no.~5, pp.
  3064--3076, May. 2020.

\bibitem{physical_layer}
T.~O’Shea and J.~Hoydis, ``{An Introduction to Deep Learning for the Physical
  Layer},'' \emph{IEEE Trans. Cogn. Commun. Netw.}, vol.~3, no.~4, pp.
  563--575, Dec. 2017.

\bibitem{wgan-gp}
F.~Zhu, X.~Wang, C.~Huang, A.~Alhammadi, H.~Chen, Z.~Zhang, C.~Yuen, and
  M.~Debbah, ``{Beamforming Inferring by Conditional WGAN-GP for Holographic
  Antenna Arrays},'' \emph{IEEE Wirel. Commun. Lett.}, pp. 1--1, May. 2024.

\bibitem{tianlin}
T.~Lin and Y.~Zhu, ``{Beamforming Design for Large-Scale Antenna Arrays Using
  Deep Learning},'' \emph{IEEE Wirel. Commun. Lett.}, vol.~9, no.~1, pp.
  103--107, Jan. 2020.

\bibitem{universal}
J.~Kim, H.~Lee, S.-E. Hong, and S.-H. Park, ``{Deep Learning Methods for
  Universal MISO Beamforming},'' \emph{IEEE Wirel. Commun. Lett.}, vol.~9,
  no.~11, pp. 1894--1898, Nov. 2020.

\bibitem{xiawenchao}
W.~Xia, G.~Zheng, Y.~Zhu, J.~Zhang, J.~Wang, and A.~P. Petropulu, ``{A Deep
  Learning Framework for Optimization of MISO Downlink Beamforming},''
  \emph{IEEE Trans. Commun.}, vol.~68, no.~3, pp. 1866--1880, Mar. 2020.

\bibitem{zhangmaojun}
M.~Zhang, J.~Gao, and C.~Zhong, ``{A Deep Learning-Based Framework for Low
  Complexity Multiuser MIMO Precoding Design},'' \emph{IEEE Trans. Wireless
  Commun.}, vol.~21, no.~12, pp. 11\,193--11\,206, Dec. 2022.

\bibitem{unfolding1}
L.~Pellaco, M.~Bengtsson, and J.~Jald{\'e}n, ``{Deep Weighted MMSE Downlink
  Beamforming},'' in \emph{ICASSP 2021 - 2021 IEEE Int. Conf. Acoust. Speech
  Signal Process. (ICASSP)}, 2021.

\bibitem{unfolding2}
Y.~Liu, Q.~Hu, Y.~Cai, G.~Yu, and G.~Y. Li, ``Deep-unfolding beamforming for
  intelligent reflecting surface assisted full-duplex systems,'' \emph{IEEE
  Trans. Wireless Commun.}, vol.~21, no.~7, pp. 4784--4800, July. 2022.

\bibitem{unfolding3}
L.~Pellaco, M.~Bengtsson, and J.~Jaldén, ``{Matrix-Inverse-Free Deep Unfolding
  of the Weighted MMSE Beamforming Algorithm},'' \emph{IEEE Open J. Commun.
  Soc.}, vol.~3, pp. 65--81, 2022.

\bibitem{unfolding4}
Q.~Hu, Y.~Cai, Q.~Shi, K.~Xu, G.~Yu, and Z.~Ding, ``{Iterative Algorithm
  Induced Deep-Unfolding Neural Networks: Precoding Design for Multiuser MIMO
  Systems},'' \emph{IEEE Trans. Wireless Commun.}, vol.~20, no.~2, pp.
  1394--1410, Feb. 2021.

\bibitem{unfolding5}
L.~Pellaco and J.~Jald{\'e}n, ``{A Matrix-Inverse-Free Implementation of the
  MU-MIMO WMMSE Beamforming Algorithm},'' \emph{IEEE Trans. Signal Process.},
  vol.~70, pp. 6360--6375, 2022.

\bibitem{zhufenghao}
F.~Zhu, B.~Wang, Z.~Yang, C.~Huang, Z.~Zhang, G.~C. Alexandropoulos, C.~Yuen,
  and M.~Debbah, ``{Robust Millimeter Beamforming via Self-Supervised Hybrid
  Deep Learning},'' in \emph{2023 31st Eur. Signal Process. Conf. (EUSIPCO)},
  2023.

\bibitem{transfer_learning_survey}
F.~Zhuang, Z.~Qi, K.~Duan, D.~Xi, Y.~Zhu, H.~Zhu, H.~Xiong, and Q.~He, ``{A
  Comprehensive Survey on Transfer Learning},'' \emph{Proc. IEEE}, vol. 109,
  no.~1, pp. 43--76, Jan. 2021.

\bibitem{MAML1}
Y.~Yuan, G.~Zheng, K.-K. Wong, B.~Ottersten, and Z.-Q. Luo, ``{Transfer
  Learning and Meta Learning-Based Fast Downlink Beamforming Adaptation},''
  \emph{IEEE Trans. Wireless Commun.}, vol.~20, no.~3, pp. 1742--1755, Mar.
  2021.

\bibitem{MAML2}
J.~Zhang, Y.~Yuan, G.~Zheng, I.~Krikidis, and K.-K. Wong, ``{Embedding
  Model-Based Fast Meta Learning for Downlink Beamforming Adaptation},''
  \emph{IEEE Trans. Wireless Commun.}, vol.~21, no.~1, pp. 149--162, Jan. 2022.

\bibitem{MLAM}
J.~Xia, S.~Li, J.-J. Huang, Z.~Yang, I.~M. Jaimoukha, and D.~Gündüz,
  ``{Metalearning-Based Alternating Minimization Algorithm for Nonconvex
  Optimization},'' \emph{IEEE Trans. Neural Netw. Learn. Syst.}, vol.~34,
  no.~9, pp. 5366--5380, Sept. 2023.

\bibitem{mlbf}
J.~Xia and D.~Gunduz, ``{Meta-learning Based Beamforming Design for MISO
  Downlink},'' in \emph{2021 IEEE Int. Symp. Inf. Theory (ISIT)}, 2021.

\bibitem{WMMSE}
Q.~Shi, M.~Razaviyayn, Z.-Q. Luo, and C.~He, ``{An Iteratively Weighted MMSE
  Approach to Distributed Sum-Utility Maximization for a MIMO Interfering
  Broadcast Channel},'' \emph{IEEE Trans. Signal Process.}, vol.~59, no.~9, pp.
  4331--4340, Sept. 2011.

\bibitem{ma2011manifold}
Y.~Ma and Y.~Fu, \emph{Manifold learning theory and applications}.\hskip 1em
  plus 0.5em minus 0.4em\relax CRC press, 2011.

\bibitem{izenman2012introduction}
A.~J. Izenman, ``{Introduction to manifold learning},'' \emph{Wiley
  Interdisciplinary Reviews: Computational Statistics}, vol.~4, no.~5, pp.
  439--446, 2012.

\bibitem{location}
C.~Feng, S.~Valaee, and Z.~Tan, ``Localization of wireless sensors using
  compressive sensing for manifold learning,'' in \emph{2009 IEEE Int. Symp.
  Pers. Indoor Mob. Radio Commun. (PIMRC)}, 2009.

\bibitem{manifoldMIMO}
X.~Zhou, P.~Wang, Z.~Yang, L.~Tong, Y.~Wang, C.~Yang, N.~Xiong, and H.~Gao,
  ``{A Manifold Learning Two-Tier Beamforming Scheme Optimizes Resource
  Management in Massive MIMO Networks},'' \emph{IEEE Access}, vol.~8, pp.
  22\,976--22\,987, 2020.

\bibitem{rethinking}
X.~Zhao, S.~Lu, Q.~Shi, and Z.-Q. Luo, ``{Rethinking WMMSE: Can Its Complexity
  Scale Linearly With the Number of BS Antennas?}'' \emph{IEEE Trans. Signal
  Process.}, vol.~71, pp. 433--446, 2023.

\bibitem{wang2023energyefficient}
X.~Wang, F.~Zhu, Q.~Zhou, Q.~Yu, C.~Huang, A.~Alhammadi, Z.~Zhang, C.~Yuen, and
  M.~Debbah, ``{Energy-efficient Beamforming for RISs-aided Communications:
  Gradient Based Meta Learning},'' in \emph{Proc. 2024 IEEE Int. Conf. Commun.
  (ICC)}, Jun. 2024.

\bibitem{sourcecode}
F.~Zhu and X.~Wang, ``{GMML},''
  \textit{\url{https://github.com/fenghaozhu/GMML}}, 2023.

\bibitem{STAR-RIS1}
X.~Mu, Y.~Liu, L.~Guo, J.~Lin, and R.~Schober, ``{Simultaneously Transmitting
  and Reflecting (STAR) RIS Aided Wireless Communications},'' \emph{IEEE Trans.
  Wireless Commun.}, vol.~21, no.~5, pp. 3083--3098, May. 2022.

\bibitem{STAR-RIS2}
M.~Ahmed, A.~Wahid, S.~S. Laique, W.~U. Khan, A.~Ihsan, F.~Xu, S.~Chatzinotas,
  and Z.~Han, ``{A Survey on STAR-RIS: Use Cases, Recent Advances, and Future
  Research Challenges},'' \emph{IEEE Internet Things J.}, vol.~10, no.~16, pp.
  14\,689--14\,711, Aug. 2023.

\bibitem{STAR-RIS3}
J.~Zuo, Y.~Liu, Z.~Ding, L.~Song, and H.~V. Poor, ``{Joint Design for
  Simultaneously Transmitting and Reflecting (STAR) RIS Assisted NOMA
  Systems},'' \emph{IEEE Trans. Wireless Commun.}, vol.~22, no.~1, pp.
  611--626, Jan. 2023.

\bibitem{compressed}
A.~Abdallah, A.~Celik, M.~M. Mansour, and A.~M. Eltawil, ``{RIS-Aided mmWave
  MIMO Channel Estimation Using Deep Learning and Compressive Sensing},''
  \emph{IEEE Trans. Wireless Commun.}, vol.~22, no.~5, pp. 3503--3521, May.
  2023.

\bibitem{finn2017model}
C.~Finn, P.~Abbeel, and S.~Levine, ``{Model-agnostic meta-learning for fast
  adaptation of deep networks},'' in \emph{Int. Conf. Mach. Learn.
  (ICML)}.\hskip 1em plus 0.5em minus 0.4em\relax PMLR, 2017.

\bibitem{3gpp.36.814}
3GPP, ``{Further Advancements for E-UTRA Physical Layer Aspects (Release 9)},''
  TS 36.814 V9.2.0, Mar. 2010.

\bibitem{xu2021robust}
W.~Xu, L.~Gan, and C.~Huang, ``{A Robust Deep Learning-Based Beamforming Design
  for RIS-Assisted Multiuser MISO Communications With Practical Constraints},''
  \emph{IEEE Trans. Cogn. Commun.}, vol.~8, no.~2, pp. 694--706, Jun. 2021.

\bibitem{3gpp.36.885}
{3GPP}, ``{Technical Specification Group Radio Access Network: Study LTE-Based
  V2X Services: (Release 14)},'' TS 36.885 V2.0.0, Jun. 2016.

\end{thebibliography}
\vspace{12pt}

\begin{IEEEbiography}[{\includegraphics[width=1in,height=1.25in,clip,keepaspectratio]{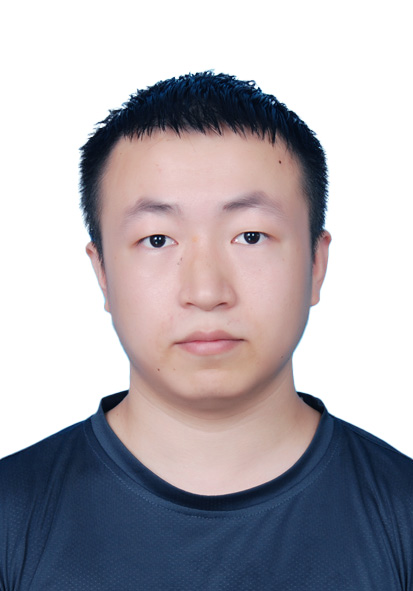}}]{Fenghao Zhu}
received the B.Eng. degree in information engineering from Zhejiang University, Hangzhou, China, in 2023, and he is currently pursuing the M.S. degree with the College of Information Science and Electronic Engineering, Zhejiang University. His current research interests include massive MIMO, signal processing, and machine learning.
\end{IEEEbiography}

\begin{IEEEbiography}[{\includegraphics[width=1in,height=1.25in,clip,keepaspectratio]{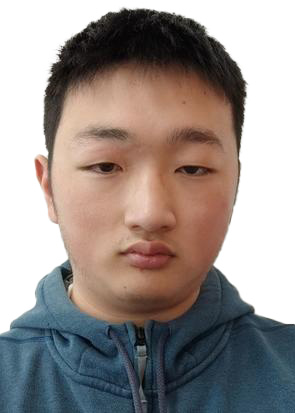}}]{Xinquan Wang}
(IEEE Student Member) is currently pursuing the B.Eng. degree at Zhejiang University, Hangzhou, China. His current research interests include 6G, beamforming and machine learning.
Mr. Wang is a recipient of 2024 IEEE ComSoc Student Grant.
\end{IEEEbiography}

\begin{IEEEbiography}[{\includegraphics[width=1in,height=1.25in,clip,keepaspectratio]{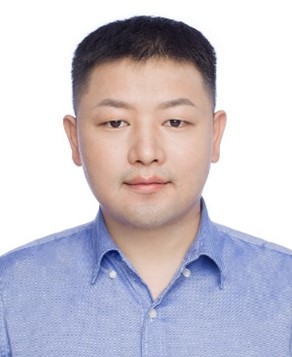}}]{Chongwen Huang} (IEEE Member)
obtained his B. Sc. degree in 2010 from Nankai University, and the M.Sc degree from the University of Electronic Science and Technology of China in 2013, and PhD degree from Singapore University of Technology and Design (SUTD) in 2019. From Oct. 2019 to Sep. 2020, he is a Postdoc in SUTD.  Since Sep. 2020, he joined into Zhejiang University as a tenure-track young professor. Dr. Huang is the recipient of 2021 IEEE Marconi Prize Paper Award, 2023 IEEE Fred W. Ellersick Prize Paper Award and 2021 IEEE ComSoc Asia-Pacific Outstanding Young Researcher Award. He has served as an Editor of IEEE Communications Letter, Elsevier Signal Processing, EURASIP Journal on Wireless Communications and Networking and Physical Communication since 2021. His main research interests are focused on Holographic MIMO Surface/Reconfigurable Intelligent Surface, B5G/6G Wireless Communications, mmWave/THz Communications, Deep Learning technologies for Wireless communications, etc.
\end{IEEEbiography}

\begin{IEEEbiography}[{\includegraphics[width=1in,height=1.25in,clip,keepaspectratio]{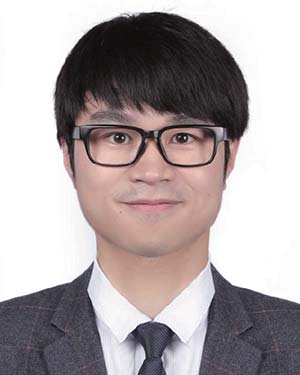}}]{Zhaohui Yang} (IEEE Member)
received the Ph.D. degree from Southeast University, Nanjing, China, in 2018. From 2018 to 2020, he was a Postdoctoral Research Associate with the Center for Telecommunications Research, Department of Informatics, King’s College London, Lodnon, U.K. From 2020 to 2022, he was a Research Fellow with the Department of Electronic and Electrical Engineering, University College London, London. He is currently a ZJU
Young Professor with the Zhejiang Key Laboratory of Information Processing Communication and Networking, College of Information Science and Electronic Engineering, Zhejiang University, Hangzhou, China. His research interests include joint communication, sensing, and computation, federated learning, and semantic communication. He was the recipient of 2023 IEEE Marconi Prize Paper Award, 2023 IEEE Katherine Johnson Young Author Paper Award, 2023 IEEE ICCCN best paper award. He was the Co-Chair for international workshops with more than ten times including IEEE ICC, IEEE GLOBECOM, IEEE WCNC, IEEE TGCN, IEEE CL, IEEE TMLCN. He has been the Guest Editor of several journals including IEEE JOURNAL ON SELECTED AREAS IN COMMUNICATIONS.
\end{IEEEbiography}

\begin{IEEEbiography}[{\includegraphics[width=1in,height=1.25in,clip,keepaspectratio]{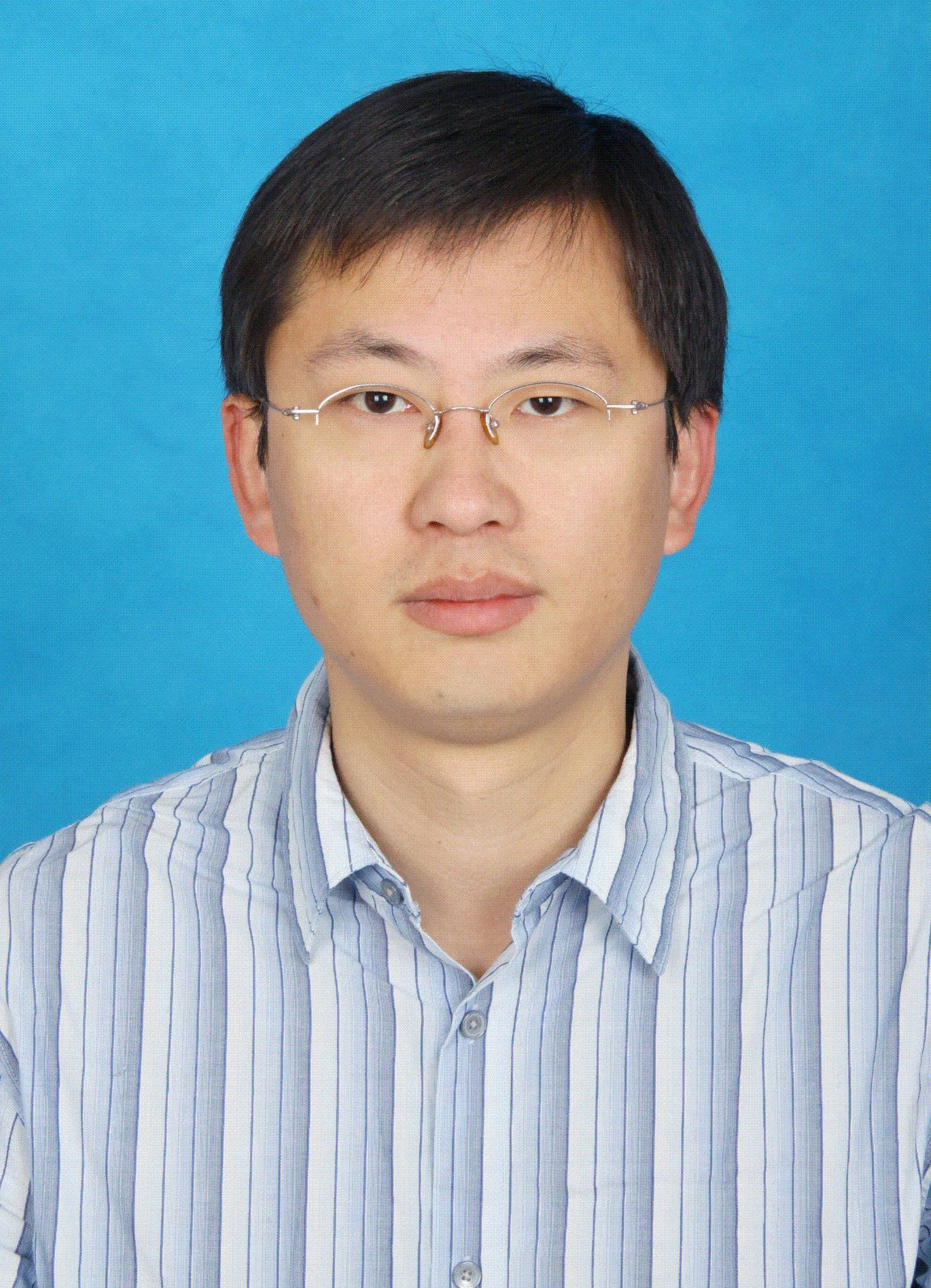}}]{Xiaoming Chen} (IEEE Member)
received the B.Sc. degree from Hohai University in 2005, the M.Sc. degree from Nanjing University of Science and Technology in 2007 and the Ph. D. degree from Zhejiang University in 2011, all in electronic engineering. He is currently a Professor with the College of Information Science and Electronic Engineering, Zhejiang University, Hangzhou, China. From March 2011 to October 2016, He was with Nanjing University of Aeronautics and Astronautics, Nanjing, China. From February 2015 to June 2016, he was a Humboldt Research Fellow at the Institute for Digital Communications, Friedrich-Alexander-University Erlangen-N\"urnberg (FAU), Germany. His research interests mainly focus on LEO satellite constellation, Internet of Things, and smart communications.
\par
Dr. Chen served as an Editor for the \textsc{IEEE Transactions on Communications} and the \textsc{IEEE Communications Letters}, and a Guest Editor for the \textsc{IEEE Journal on Selected Areas in Communications} ``Massive Access for 5G and Beyond" and the \textsc{IEEE Wireless Communications} ``Massive Machine-Type Communications for IoT". He received the Best Paper Awards at the IEEE Global Communications Conference (GLOBECOM) 2020, the International Conference on Wireless Communications and Signal Processing (WCSP) 2020, the IEEE International Conference on Communications (ICC) 2019, and the IEEE/CIC International Conference on Communications in China (ICCC) 2018.
\end{IEEEbiography}

\begin{IEEEbiography}[{\includegraphics[width=1in,height=1.25in,clip,keepaspectratio]{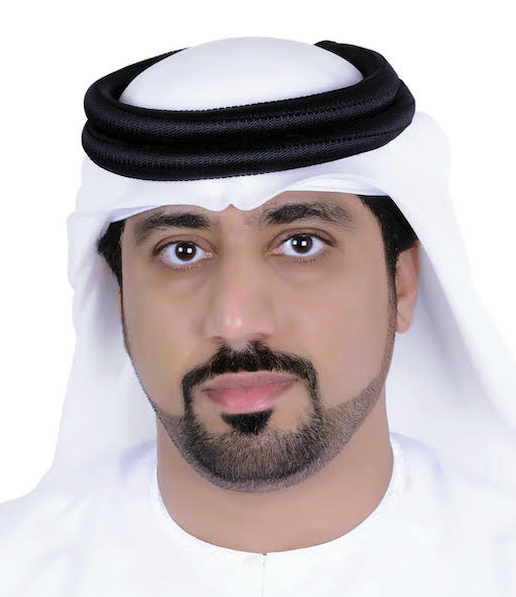}}]{Ahmed Al Hammadi} (IEEE Member)
is currently a Lead Researcher at the Artificial Intelligence Cross Center Unit, part of the Technology Innovation Institute in Abu Dhabi, UAE. He earned his PhD in Electrical and Computer Engineering in 2022, his Master’s in Communications Engineering in 2015, and his Bachelor’s in Electronics Engineering in 2011, all from Khalifa University. Ahmed began his career in research and development immediately after completing his undergraduate studies in 2011.
\par
Over the years, Ahmed has increasingly focused on applying Artificial Intelligence (AI) techniques to optimize wireless networks. This has led him to delve deeper into broader AI and Machine Learning (ML) fields. Ahmed's work now encompasses the development of innovative AI and ML solutions to address complex problems across various domains.
\par
In 2024, Ahmed achieved first place in the Global Signal Processing Grand Challenge organized by MIT during the ICASSP conference, using an innovative deep neural network (DNN) based solution. Additionally, in 2011, he was selected as one of 60 elite Emiratis for the prestigious Al Nokhba internship at Global Foundries. Ahmed is actively engaged in early research towards future technological advancements and has demonstrated his commitment to the field through several patents, contributions to key book chapters, and multiple high-impact presentations at international conferences. His passion for advancing technologies through AI and ML drives him to continually explore and implement new innovations in this dynamic and ever-evolving field.
\end{IEEEbiography}

\begin{IEEEbiography}[{\includegraphics[width=1in,height=1.25in,clip,keepaspectratio]{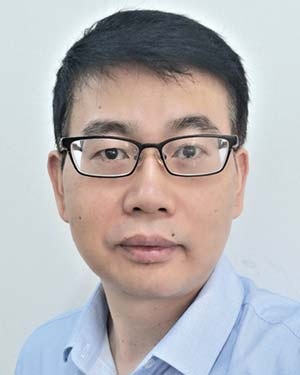}}]{Zhaoyang Zhang} (IEEE Senior Member)
received the Ph.D. degree from Zhejiang University, Hangzhou, China, in 1998, where he is currently a Qiushi Distinguished Professor. He has authored or coauthored more than 300 peer-reviewed international journal and conference papers, including eight conference best papers awarded by IEEE ICC 2019 and IEEE GlobeCom 2020. His research interests include the fundamental aspects of wireless communications
and networking, such as information theory and coding theory, network signal processing and distributed learning, AI-empowered communications and networking, synergetic sensing, and computing and communication. Dr. Zhang has been the Editor of IEEE TRANSACTIONS ON WIRELESS COMMUNICATIONS, IEEE TRANSACTIONS ON COMMUNICATIONS, and IET Communications, and the General Chair, TPC Co-Chair or Symposium Co-Chair for PIMRC 2021 Workshop on Native AI Empowered Wireless Networks, VTC-Spring 2017 Workshop on HMWC, WCSP 2013/2018, and Globecom 2014 Wireless Communications Symposium. He was also a keynote speaker for APCC 2018 and VTC-Fall 2017 Workshop NOMA. He was the recipient of the National Natural Science Fund for Distinguished Young Scholars by NSFC in 2017.
\end{IEEEbiography}

\begin{IEEEbiography}[{\includegraphics[width=1in,height=1.25in,clip,keepaspectratio]{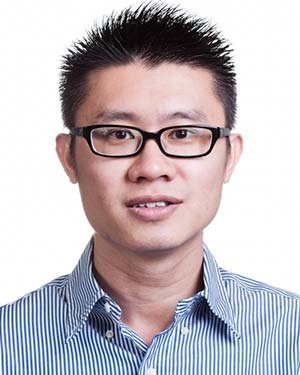}}]{Chau Yuen} (Fellow, IEEE)
received the B.Eng. and Ph.D. degrees from Nanyang Technological University, Singapore, in 2000 and 2004, respectively. In 2005, he was a Postdoctoral Fellow with Lucent Technologies Bell Labs, Murray Hill, NJ, USA. From 2006 to 2010, he was with the Institute for Infocomm Research, Singapore. From 2010 to 2023, he was with the Engineering Product Development Pillar, Singapore University of Technology and Design, Singapore. Since 2023, he has been with the School of Electrical and Electronic Engineering, Nanyang Technological University. He has three U.S. Patents and authored or coauthored more than 400 research papers at international journals. Dr. Yuen was the recipient of IEEE Communications Society Leonard G. Abraham Prize (2024), IEEE Communications Society Best Tutorial Paper Award (2024), IEEE Communications Society Fred W. Ellersick Prize (2023), IEEE Marconi Prize Paper Award in Wireless Communications (2021), IEEE APB Outstanding Paper Award (2023), and EURASIP Best Paper Award for Journal on Wireless Communications and Networking (2021). He has been the Editor-in-Chief of Springer Nature Computer Science, Editor of IEEE TRANSACTIONS ON VEHICULAR TECHNOLOGY, IEEE SYSTEMS JOURNAL, and IEEE TRANSACTIONS ON NETWORK SCIENCE AND ENGINEERING, where he was the recipient of IEEE TNSE Excellent Editor Award and Top Associate Editor for TVT from 2009 to 2015. He was the Guest Editor of several special issues, including IEEE JOURNAL ON SELECTED AREAS
IN COMMUNICATIONS, IEEE WIRELESS COMMUNICATIONS, IEEE COMMUNICATIONS MAGAZINE, IEEE VEHICULAR TECHNOLOGY MAGAZINE, IEEE TRANSACTIONS ON COGNITIVE COMMUNICATIONS AND NETWORKING, and Elsevier Applied Energy. He is a Distinguished Lecturer of IEEE Vehicular Technology Society, Top 2\% Scientists by Stanford University, and also a Highly Cited Researcher by Clarivate Web of Science.
\end{IEEEbiography}

\begin{IEEEbiography}[{\includegraphics[width=1in,height=1.25in,clip,keepaspectratio]{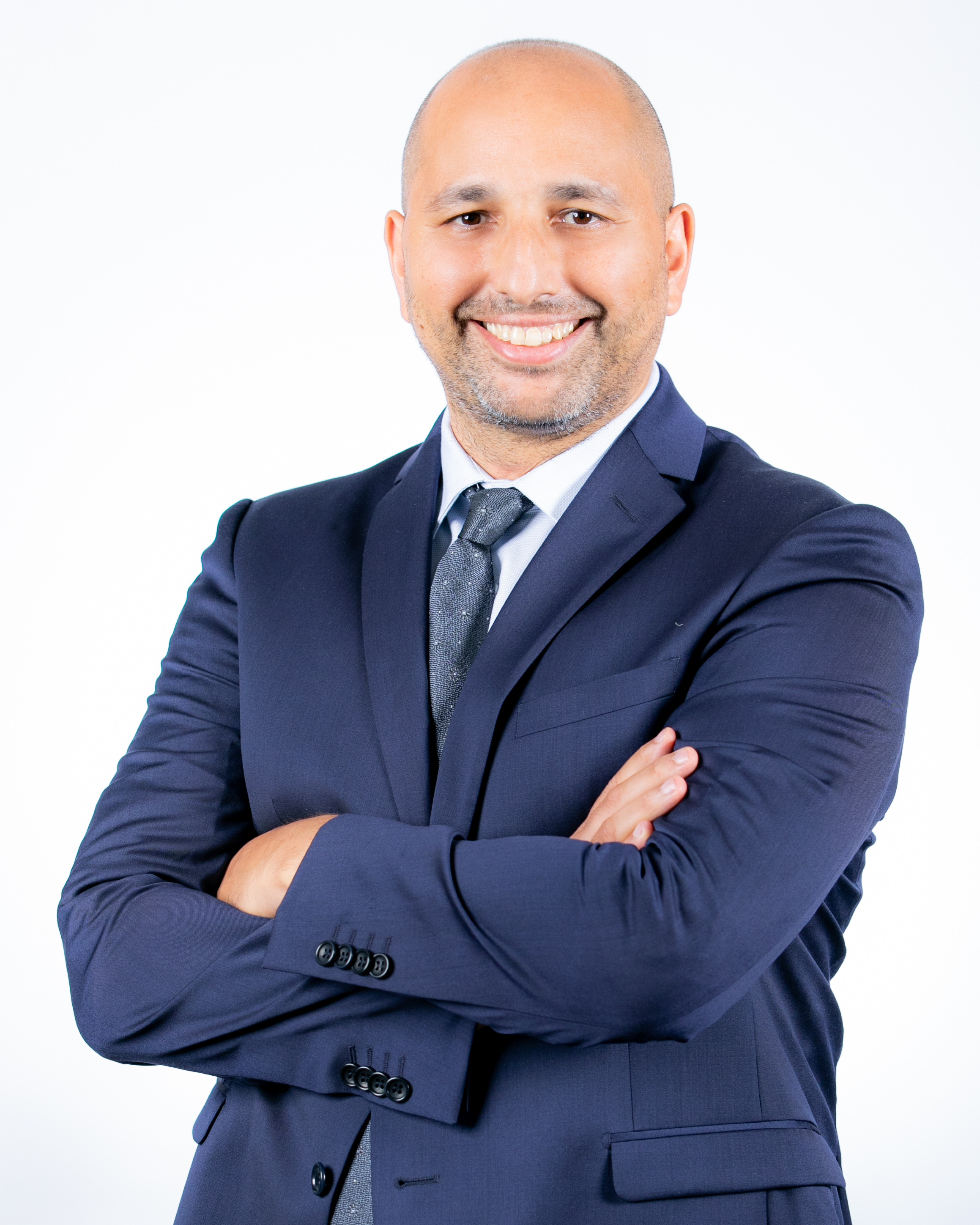}}]{M\'{e}rouane~Debbah} (Fellow, IEEE)
is Professor at Khalifa University of Science and Technology in Abu Dhabi and founding Director of the KU 6G Research Center. He is a frequent keynote speaker at international events in the field of telecommunication and AI. His research has been lying at the interface of fundamental mathematics, algorithms, statistics, information and communication sciences with a special focus on random matrix theory and learning algorithms. In the Communication field, he has been at the heart of the development of small cells (4G), Massive MIMO (5G) and Large Intelligent Surfaces (6G) technologies. In the AI field, he is known for his work on Large Language Models, distributed AI systems for networks and semantic communications. He received multiple prestigious distinctions, prizes and best paper awards for his contributions to both fields. He is an IEEE Fellow, a WWRF Fellow, a Eurasip Fellow, an AAIA Fellow, an Institut Louis Bachelier Fellow and a Membre émérite SEE.
\end{IEEEbiography}

\end{document}